\documentclass[12pt]{article}
\pdfoutput=1
\usepackage{jheppub}
\usepackage[utf8]{inputenc}
\usepackage{verbatim}
\usepackage{amsmath}
\usepackage{amssymb}
\usepackage{amsthm}
\usepackage{amsfonts}
\usepackage{hyperref}
\usepackage{MnSymbol}
\usepackage{xcolor}
\newcommand{\be}{\begin{equation}}
\newcommand{\ee}{\end{equation}}
\newcommand{\ran}{\rangle}
\newcommand{\lan}{\langle}

\newcommand{\ol}{\overline}
\newcommand{\wt}{\widetilde}
\newcommand{\bi}{\begin{itemize}}
\newcommand{\ei}{\end{itemize}}
\newcommand{\Tr}{\mathrm{Tr}}
\newcommand{\Hh}{\mathcal{H}}
\newcommand{\M}{\mathcal{M}}
\newcommand{\bfig}{\begin{figure}\begin{center}}
\newcommand{\efig}{\end{center}\end{figure}}

\newcommand{\T}{\mathcal{T}}

\newcommand{\J}{\mathcal{J}}

\usepackage{graphicx}
\begin{document}
\title{Observers, $\alpha$-parameters, and the Hartle-Hawking state}
\author[a]{Daniel Harlow}
\affiliation[a]{Center for Theoretical Physics - a Leinweber Institute\\ Massachusetts Institute of Technology, Cambridge, MA 02139, USA}
\emailAdd{harlow@mit.edu}
\abstract{In this paper we extend recent ideas about observers and closed universes to theories where observers can be fluctuated into existence in the Hartle-Hawking state.  
This introduces a phenomenon that was not considered in these earlier discussions: the dominant transition from one cosmological state to another can go through a fluctuation that annihilates the universe and creates a new one.  We nonetheless argue that the observer decoherence rule of \cite{Harlow:2025pvj} allows for the third-quantized description of such a theory to emerge from a factorizing holographic theory with a one-dimensional Hilbert space, without any need for $\alpha$-parameters.  We also point out a close analogy between the observer rule in this context and the coarse-graining of the spectral form factor at late times for AdS black holes. Along the way we clarify several aspects of the relationship between holography, the gravitational path integral, and $\alpha$-parameters. We also explain why string theory scattering amplitudes do not lead to a one-dimensional Hilbert space on the worldsheet, despite being computed by a gravitational path integral with a sum over topology.  Finally we point out that using the path integral to compute integrated local operators conditioned on an observer in the context of a theory with a landscape can lead to rather surprising conclusions.  For example we argue that in a landscape with one AdS minimum and one dS minimum, both of which can support observers, an observer almost surely finds themself in dS and not AdS even if the boundary conditions are dual to a state with an observer in AdS.}
\maketitle

\section{Introduction}
In recent years there has been substantial progress on the black hole information problem, leading to a fairly precise proposal for the emergence mechanism of the interior degrees of freedom from the fundamental microstates.  This proposal can be summarized as follows \cite{Penington:2019npb,Almheiri:2019psf,Penington:2019kki,Akers:2022qdl}: 
\bi
\item[] The effective field theory of the black hole interior is non-isometrically encoded into its microstate degrees of freedom, leading to a Page curve \cite{Page:1993df,Page:2013dx} that is consistent with unitarity via the quantum extremal surface formula \cite{Engelhardt:2014gca}.  The failure of isometry cannot be detected by operations whose complexity is subexponential in the entropy of the black hole.  
\ei
One of the most compelling sources of evidence for this proposal is the gravitational path integral, where the sum over topologies naturally includes contributions that connect the different copies of the state of the system in entropy calculations even though such contributions do not appear when one uses the path integral to compute the state itself \cite{Lewkowycz:2013nqa}.  This seeming mismatch is explained by the idea that the gravitational path integral only captures features of the fundamental theory which are coarse-grained or averaged in some way \cite{Cotler:2016fpe,Saad:2019lba}.  Including these topologies gives answers which are consistent with the above proposal, in particular including a unitary Page curve and the ability to reconstruct interior operators on the Hawking radiation \cite{Penington:2019kki,Almheiri:2019qdq,Marolf:2020xie,Marolf:2020rpm}.  These path integral contributions can be derived explicitly in toy models that concretely implement the proposal \cite{Akers:2022qdl}.  

There remains much still to understand about this proposal.  What exactly goes wrong when exponential operations are performed?  Is the breakdown of the proposal for generic microstates really signaling that these do not have a semiclassical interior?  Can the proposal be derived from some fundamental theory of quantum gravity such as string theory?  Can it be realized concretely in models that capture more features of the real world?  Work continues on answering these questions, and indeed some progress has been made \cite{Stanford:2022fdt,Bousso:2023kdj,Boruch:2024kvv,Almheiri:2025ugo}, but it has also proven irresistible to push forward and see what the ideas behind the proposal might tell us about the \textit{other} big problem in quantum gravity: how to think about the quantum mechanics of the universe as a whole?  

Attempts to generalize our improved understanding of black holes to cosmology almost immediately run into a seemingly insurmountable problem \cite{Almheiri:2019hni,Penington:2019kki,McNamara:2020uza,Usatyuk:2024mzs,Usatyuk:2024isz} : 
\bi
\item[] The Hilbert space of quantum gravity in a closed universe has dimension one.  In other words the total number of degrees of freedom in the universe is zero.
\ei
There are various arguments for this conclusion, see section 2 of \cite{Harlow:2025pvj} for a review.  Perhaps the simplest (although also the least precise) is that holography says that in quantum gravity the fundamental degrees of freedom live at the spatial boundary, so if there is no spatial boundary then there are no fundamental degrees of freedom.  What are we to make of this?  One possibility is to simply conclude that we do not live in a closed universe.  This may be correct, but it would be rather surprising if we could infer global properties of the universe using local observations.  Another possibility is that the arguments are all wrong, but then we would need to find the mistake and explain why it doesn't invalidate all the recent progress on black holes.  A final possibility is that the universe indeed has zero degrees of freedom, but that our daily experiences are somehow consistent with this.  In \cite{Abdalla:2025gzn,Harlow:2025pvj} it was proposed that approximate semiclassical physics in a closed universe can be recovered by modifying the laws of physics to treat the observer differently from the rest of the matter in the universe (for earlier antecedents to this work see \cite{Dong:2020uxp,Antonini:2023hdh,Sahu:2024ccg}, and for the idea of implementing an observer as a form of gauge-fixing see \cite{Chandrasekaran:2022cip,Witten:2023xze}).  One way to think about these proposals is that they implement Bohr's old idea that quantum mechanics must be understood as a theory of a classical observer interacting with a quantum system.  See \cite{Akers:2025ahe,Chen:2025fwp,Blommaert:2025bgd,Engelhardt:2025azi,Antonini:2025ioh,Higginbotham:2025clp} for some further discussion of these proposals, as well as \cite{Antonini:2024mci,Engelhardt:2025vsp,Gesteau:2025obm,Liu:2025cml,Kudler-Flam:2025cki} for more related discussion.   

Although there is some evidence that the proposals of \cite{Abdalla:2025gzn,Harlow:2025pvj} may give a way to have a semiclassical gravity theory emerge from a holographic theory of a closed universe, it must be acknowledged that they are still rather speculative.  The goal of this paper is to address several questions about closed universes and observers that have come up over the last year: 
\bi
\item[(1)] How does the one state of holography in a closed universe relate to the Hartle-Hawking (HH) state \cite{Hartle:1983ai}?  Do they even live in the same Hilbert space?
\item[(2)] In the third-quantized approach to quantum cosmology introduced in \cite{Coleman:1988cy,Giddings:1988cx,Coleman:1988tj}, and recently advocated in \cite{Marolf:2020xie}, which we will refer to as \textbf{Baby Universe Field Theory (BUFT)}, there is a large closed-universe Hilbert space spanned by a set of mutual orthogonal ``$\alpha$-states''.  How is this related to the one-state claim, and in particular does it give a way out of it?
\item[(3)] In the perturbative approach to closed string theory we sum over splitting and joining worldsheets, but the Hilbert space of closed string states is not one-dimensional.  How is this compatible with the one-state claim?
\item[(4)] What is the relationship, if any, between the observer proposal and BUFT?  Are there really two ways to get a large closed universe Hilbert space?
\ei
In answering these questions a key issue that will arise is to what extent the gravitational path integral is well-defined.  As we will review below, to complete the program of BUFT and construct well-defined $\alpha$-states it is necessary to have a renormalizable theory of gravity with a convergent sum over topology.  As far as we know this is only possible in $1+1$ dimensions.\footnote{A possible higher-dimensional example is pure gravity in $2+1$ dimensions, which is also renormalizable, but the sum over topologies is currently under rather poor control.  This may not be a coincidence: the dual of $AdS_3$ should be $CFT_2$, but as we will review below conformal field theories do not allow $\alpha$-parameters.}  When it is possible, the answer to questions (1) and (2) is that each $\alpha$-state is the one state for a holographic theory, while the HH state is a coherent superposition over many $\alpha$-states \cite{Marolf:2020xie}.  Moreover thinking of the perturbative string worldsheet in this way we will see that it again has a large Hilbert space dimension because there are many $\alpha$-states. 

In higher dimensions however the gravitational path integral is too ambiguous to allow construction of well-defined  $\alpha$-states.  As we will discuss in section \ref{alphasec} this is probably a good thing, since $\alpha$-parameters are not compatible with string theory or holography \cite{McNamara:2020uza}.  On the other hand we could still hope to get some kind of BUFT as an emergent notion, starting from some precise holographic theory.  How then are we supposed to think about questions (1-4) in such a context?  We will study this in a concrete code model of holography in a closed universe, leading to the following proposed answers: 
\bi
\item[(1)]  The machinery of holographic codes allows us to view the ``one state'' of holography as living in BUFT Hilbert space.  It is not however equal to the HH state, as the former is strongly UV-sensitive while the latter can be computed using low-energy methods.\footnote{The relationship between the two states is somewhat similar to the relationship between the final pure state of an evaporating black hole and the Hawking entangled state of interior and exterior modes.}  In particular the HH state  is only one of many in the emergent description.  
\item[(2)] In a holographic theory there are no $\alpha$-parameters to consider, so there is no large Hilbert space spanned by $\alpha$-states.    
\item[(3)] The closed string worldsheet is not a holographic theory, so it does not need to have one state.  In string theory holography operates in the target space.
\item[(4)] BUFT can approximately emerge from a fixed holographic theory.  In general this relies on adopting some version of the observer proposal of \cite{Harlow:2025pvj}, although in some cases the observer is only really needed for gauge fixing as in \cite{Chandrasekaran:2022cip,Witten:2023xze}.  The observer rule of \cite{Abdalla:2025gzn} does \textit{not} always work, as it discards contributions to the path integral that are there in BUFT and sometimes dominant.  
\ei
We will also point out that there is a natural analogy between the observer proposal of \cite{Harlow:2025pvj} and the coarse-grained discussion of the spectral form factor in \cite{Cotler:2016fpe}, where in both cases some averaging over boundary sources is necessary to realize the emergence of the gravitational path integral in a fixed holographic theory.

\bfig
\includegraphics[height=5cm]{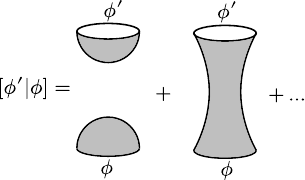}
\caption{The closed universe inner product in baby universe field theory.  The disconnected contribution is not present in the canonical gravity calculation of this inner product, so when it is big it gives a large deviation from quantum field theory in a fixed background even when $G_N$ is small.}\label{buftipfig}
\efig
Questions (1-4) address the relationship between BUFT and a fundamental holographic theory.  There is also the interesting question of how BUFT relates to the canonical approach to quantum gravity.  This is important because it is canonical quantum gravity (CQG) that reduces to quantum field theory in curved space in the limit $G_N\to 0$ of vanishing Newton constant.  In general however BUFT does not reduce to CQG when $G_N$ is small.  The problem is shown in figure \ref{buftipfig}: the BUFT transition amplitude between two closed universe states is sometimes dominated by a disconnected contribution where the initial state of the universe ``unfluctuates'' through the Hartle-Hawking state and then ``refluctuates'' in the final state.  Such contributions do not arise in CQG, where the inner product between the two states on a spatial manifold $\Sigma$ is computed using the path integral on the cylinder topology $[0,1]\times \Sigma$.  When these disconnected contributions are large compared to the cylinder they cause a version of the infamous ``Boltzmann brain'' problem: the dominant way to form structure is by fluctuation,  so the state that we prepare the system in does not matter for later observations.  In other words initial states which naively are distinguishable are in fact almost the same. 

The question of when the inner product is dominated by the cylinder is a subtle one, and we will not give a systematic study.  In particular in AdS it is rather dependent on details of the states, see for example \cite{Marolf:2021kjc} (our expectation however is that in states which evolve to nice Lorentzian bang/crunch universes the cylinder dominates).  The most interesting case where the disconnected contribution definitely dominates is de Sitter space, which indeed has long been understood to have a Boltzmann brain problem \cite{Goheer:2002vf,Carroll:2017gkl}.   

It is perhaps the case that we should simply discard states where the inner product is dominated by disconnected geometries on phenomenological grounds.  It is worth considering however if there is still some way to do computations in such states which allows for field theory in curved space to emerge at least for local experiments.  In section \ref{patchsec} we describe a potential  way to do this via the introduction of \textbf{patch operators}, which are local operators dressed to an observer.  We will see that when the sphere partition function is large this gives a fairly satisfactory way to reproduce canonical answers.

It is interesting to consider what the ideas in this paper have to say about a quantum gravity theory with a landscape of metastable vacua, such as seems to exist in string theory \cite{Susskind:2003kw}.  This is a hard problem of course, and one that we will of course not fully address, but in section \ref{patchsec} we use patch operators to do some preliminary calculation.  What we will argue that in a landscape with an AdS vacuum and a dS vacuum, if we prepare the theory in a state which we conventionally would say describes an observer in AdS then actually the most likely place to find an observer is in dS in the HH state.  In other words in such a theory the bulk dual of a fairly standard CFT state is afflicted by the Boltzmann brain problem!  Clearly this needs to be considered further, since for example $\mathcal{N}=4$ super Yang-Mills theory is indeed expected to be dual to a quantum gravity theory (IIB string theory) that plausibly has metastable dS vacua \cite{Kachru:2003aw,McAllister:2024lnt}.  

\textbf{Note:} This paper is the result of collaboration and discussion with Ying Zhao.  In the end we had different opinions about what to emphasize, so we are each writing our own paper.  We are in agreement about technical results.  

\subsection{A comment on notation}
In this paper there are three kinds of quantum states, which are denoted $|\psi\ran$, $|\psi]$, and $|\psi\rrangle$.  The first is a state in the Hilbert space of canonical quantum gravity with fixed spatial topology, the second is a state in the third-quantized Hilbert space of baby universe field theory, and the third is a number in the one-dimensional Hilbert space of holographic quantum gravity in a closed universe.  In low spacetime dimensions all three things are related via the magic of averaging over $\alpha$-parameters, but they are not the same.  In higher dimensions there are no $\alpha$-parameters, and it is only the holographic description that is well-defined.  The logic can be subtle at times, so paying careful attention to which kind of state is appearing in each quantity is essential.

\section{Gravitational path integral for closed universes}\label{pisec}
We begin by reviewing the gravitational path integral approach to quantum cosmology.  There are (at least) three distinct ways of interpreting this path integral, which we will refer to as canonical quantum gravity (CQG), baby universe field theory (BUFT), and averaged holography (AH), and we will discuss each in turn.  In this section we will assume that the path integral is well-defined, meaning that we will ignore the issues of non-renormalizability and convergence of the sum over topology.  We will return to these issues in section \ref{alphasec} below.  

\subsection{Canonical quantum gravity}
The most conventional approach to closed universe quantum gravity is canonical quantization.  For each connected spatial topology $\Sigma$ we have a pre-Hilbert space $\mathcal{H}_\Sigma$ of functionals $\Psi[\phi]$, where $\phi$ are the dynamical fields evaluated on $\Sigma$.  The physical Hilbert space is then constructed as the set of diffeomorphism-invariant states.\footnote{\label{gafoot}There is some subtlety in passing to the physical Hilbert space due to the noncompact nature of the diffeomorphism group; one has to use some kind of group-averaging to define a physical inner product on invariant states \cite{Marolf:2000iq,Held:2025mai,Alonso-Monsalve:2025lvt}.}  In the path integral approach to canonical quantization we have the expression
\be\label{phiip}
\lan\phi'|P|\phi\ran=\int \mathcal{D}\phi|_{\phi}^{\phi'} e^{iS[\phi,\Sigma\times I]},
\ee
where $P$ is the projection onto diffeomorphism-invariant states and $S[\phi,\Sigma\times I]$ is the Lorentzian action evaluated on the Lorentzian manifold $\Sigma \times I$ with $I$ a time interval.  See figure \ref{cqgipfig} for an illustration.

\bfig
\includegraphics[height=6cm]{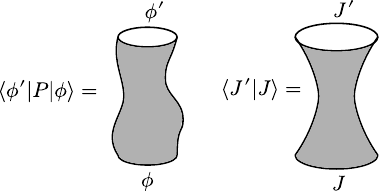}
\caption{Computing the closed universe inner product in canonical quantum gravity.  On the left the Lorentzian path integral computes the matrix elements of the projection onto diff-invariant states in the field basis as in \eqref{phiip}, while on the right we use the Euclidean path integral to compute the inner product between  states prepared by Euclidean AdS boundaries with sources $J$ an $J'$ as in \eqref{CQGIP}. }\label{cqgipfig}
\efig
The projection onto gauge-invariant states in this expression is somewhat inconvenient to keep track of, as are the group-averaging subtleties mentioned in footnote \ref{gafoot}.  In the context of closed universes with negative cosmological constant Marolf and Maxfield suggested a useful way to avoid both issues.  Namely we can consider states prepared by inserting Euclidean AdS boundaries with topology $\Sigma$ in the future and past, each carrying some set of sources $J$, and then compute an inner product using the Euclidean cylinder:
\be\label{CQGIP}
\lan J'|J\ran=\int \mathcal{D}\phi|_J^{J'}e^{-S_{E}[\phi,\Sigma\times I]},
\ee
where the action $S_{E}$ is now defined on a Euclidean cylinder.  See figure \ref{cqgipfig} for an illustration.  This way of preparing states is particularly natural when we want to consider some fundamental holographic description of the system, since these boundary sources are well-defined in AdS/CFT in a way that states on bulk Cauchy surfaces are not, so in what follows we will mostly stick to this asymptotic notation for states.  

There are several reasons why CQG is not so popular these days.  Perhaps the most decisive is that it does not predict a unitary black hole S-matrix \cite{Hawking:1976ra}, and one can also worry that it is in some sense nonlocal to not sum over all topologies that are consistent with the boundary conditions.  That said, CQG has several substantial advantages over BUFT and AH:
\bi
\item It is manifestly equal to quantum mechanics, with the path integral being derived from the operator formalism in the standard way.  This is to be compared with BUFT, which is not manifestly equal to quantum mechanics, and AH, which is almost never equal to quantum mechanics.  
\item It reduces to quantum field theory in a fixed background in the limit that $G\to 0$.  As quantum field theory in a fixed background is good enough to describe all of the nongravitational phenomena we have observed, the same had better be true for whatever theory of quantum gravity we finally adopt.  
\item One does not need to worry about the convergence of the sum over topologies, as there is no such sum.
\ei

\subsection{Baby universe field theory}
Already in the mid 1970s it was clear that there are situations where it is a good idea to sum over topologies in the gravitational path integral that do not have a canonical interpretation \cite{Gibbons:1976ue}.  This idea was further developed by Coleman, Giddings, and Strominger into the third-quantized formalism that we call BUFT \cite{Coleman:1988cy,Giddings:1988cx,Coleman:1988tj}.  This formalism was recently nicely refined in \cite{Marolf:2020xie}, whose presentation we now review.  To avoid discussion of diffeomorphism constraints we will use states prepared by Euclidean AdS boundaries, but the same idea works for field eigenstates on Cauchy surfaces.  

\subsubsection{Defining the inner product}
\bfig
\includegraphics[height=6cm]{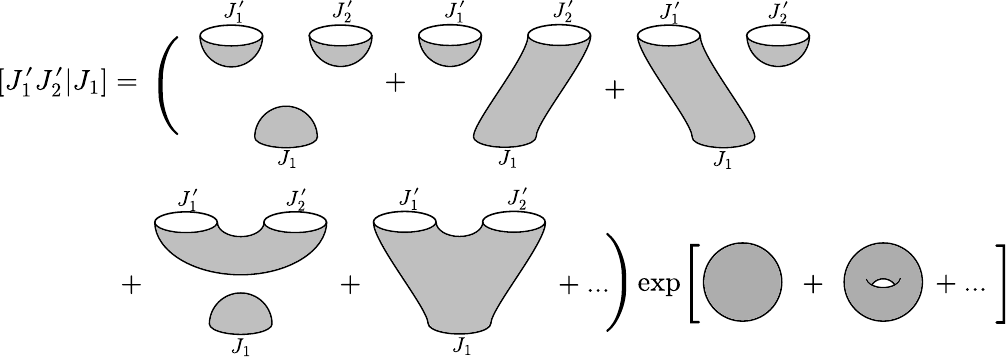}
\caption{Computing the overlap of a one-universe state and a two-universe state in BUFT.  The disconnected vacuum bubbles factor out and exponentiate.  In CQG none of these topologies would contribute and the overlap would be zero.}\label{BUFTfig}
\efig
The basic construction of BUFT is that we define a pre-Hilbert space $\Hh_{BUFT}^0$ spanned by states of the form $|J_1\ldots J_N]$, each of which describes $N$ closed universes with AdS sources $J_1 \ldots J_N$.  Each $J_i$ includes a spatial topology and induced metric, together with possible matter sources.  These universes are treated as bosons, meaning for example that $|J_1 J_2]=|J_2 J_1]$.\footnote{The reason for this is that we should treat spatial diffeomorphisms that exchange connected spatial components as gauge symmetries.  if we didn't do this the path integral would assign norm zero to the state $|J_1J_2]-|J_2J_1]$, so we anyways would get rid of it when we quotient by null states below.}  We then use the path integral to define a hermitian product
\be\label{BUFTIP}
[J'_1\ldots J_N'|J_1\ldots J_M]\equiv\sum_{\M}\int \mathcal{D}\phi|_J^{J'} e^{-S_E[\phi,\M]}
\ee
where the sum is over all Euclidean spacetime manifolds $\M$, including disconnected ones, obeying the right boundary conditions in the future and past.  See figure \ref{BUFTfig} for an illustration.  The sum over topologies ensures that this product respects boson statistics.  Note that we are using square brackets for states in the BUFT pre-Hilbert space to distinguish them from states in the CQG Hilbert space.  

As emphasized in \cite{Marolf:2020xie} the hermitian product \eqref{BUFTIP} in general is not positive-definite, or even positive-semidefinite (see also \cite{Colafranceschi:2023moh}).  This is why in the previous subsection we said that BUFT is not manifestly equal to quantum mechanics.  In theories where the hermitian product \eqref{BUFTIP} \textit{is} positive semi-definite, then we can quotient $\mathcal{H}_0$ by addition of null states to get the true Hilbert space $\Hh_{BUFT}$.  In this section we will simply assume the positive semidefiniteness of \eqref{BUFTIP}.

To make a more direct comparison between BUFT and CQG, we can do a direct sum over symmetrized tensor products of CQG Hilbert spaces with all possible spatial topologies (in CQG with disconnected space we should still gauge diffs that exchange components). The states $|J_1\ldots J_N\ran$ that span this space have the same labels as the states $|J_1\ldots J_N]$ that span $\Hh_{BUFT}^0$.  The inner product is different however, since in \eqref{BUFTIP} we sum over all topologies while in the multi-universe generalization of \eqref{CQGIP} we only sum over the ways of connecting ket and bra spatial components with the same topology using cylinders.  So for example none of the topologies in figure \ref{BUFTfig} would contribute in CQG, and indeed any overlap between states with different numbers of connected components vanishes in CQG.

It is important to mention that due to the symmetry between bras and kets in the Euclidean path integral there is a natural conjugation operation on sources, conventionally called CPT \cite{Coleman:1988cy,Marolf:2020xie} or CRT \cite{Harlow:2023hjb}, recently instead called $\T$ or $T$ depending on whether or not we have time-reversal symmetry \cite{Witten:2025ayw}, which converts a bra source into a ket source:
\be\label{conjJ}
[JJ_1'\ldots J_N'|J_1\ldots J_M]=[J_1'\ldots J_N'|J^*J_1\ldots J_M].
\ee

In BUFT there is a special state to consider, which is the Hartle-Hawking state $|HH]$.  The transition amplitude for the HH state to any other multi-universe state is defined by the path integral with no past boundaries:
\be
[J_1\ldots J_N|HH]=\sum_\M \int \mathcal{D}\phi|^Je^{-S_E[\phi,\M]}.
\ee
The norm of the Hartle-Hawking state is the sum over all manifolds with no boundaries:
\be
\aleph\equiv [HH|HH] = \sum_\M \int \mathcal{D}\phi e^{-S_E[\phi,\M]}=\exp\left[\sum_{\M \, {\rm connected}} \int \mathcal{D}\phi e^{-S_E[\phi,\M]}\right].
\ee
In the last equation we used the combinatorics familiar from Feynman diagrams to write the sum over all (possibly disconnected) geometries (including the empty set, whose action is zero) as the exponential of the sum over connected geometries.  This is the exponential factor we pulled out in figure \ref{BUFTfig}.  For general inner products between closed universe states we can factor out the disconnected diagrams, to get the nice formula
\be
\aleph^{-1}[J'_1\ldots J_N'|J_1\ldots J_M]=\sum_{\M \, {\rm connected}}\int \mathcal{D}\phi|_J^{J'} e^{-S_E[\phi,\M]}.
\ee
Here ``connected'' means that each connected component of $\M$ should contain at least one asymptotic boundary component.  For example the right-hand side here is the quantity in round parentheses in figure \ref{BUFTfig}.

It is worth briefly commenting on the relationship between the BUFT state $|HH]$ and the state originally defined by Hartle and Hawking.  The original Hartle-Hawking state was defined in the context of CQG, and in our notation it is given by
\be
\lan J|HH\ran=[J|HH].
\ee
This equation may look trivial, but it is not since since the inner products are different in CQG and BUFT.  In particular the norms of these states are not the same:\footnote{A historical remark: this distinction was important in respective attempts by Hawking and Coleman to use Euclidean gravity to explain why the cosmological constant should vanish \cite{Hawking:1984hk,Coleman:1988tj}.} 
\begin{align}\nonumber
[HH|HH]&=\aleph\\
\lan HH|HH\ran&=\log \aleph.
\end{align}
The primary role for the canonical HH state $|HH\ran$ in this paper is that we can think of it as giving the disconnected contribution to the one-universe inner product in BUFT:
\be\label{cqgHH}
\aleph^{-1}[J'|J]\supset \lan J'|HH\ran\lan HH|J\ran.
\ee

\subsubsection{Universe fields and $\alpha$-states}
One of the essential ingredients of BUFT are field operators $\hat{Z}(J)$ that create universes.  In the above notation the field operators are defined to obey
\be
\hat{Z}(J)|J_1\ldots J_N]=|JJ_1\ldots J_N],
\ee
and due to our conjugation equation \eqref{conjJ} we have\footnote{In the notation of \cite{Coleman:1988cy} we would write $\hat{Z}(J)=a_{J^*}+a_J^\dagger$, where $a_J$ is an annihilation operator for baby universes.  This makes it clear that $\hat{Z}$ is a field operator in third quantization.}
\be\label{conjZ}
\hat{Z}(J)^\dagger=Z(J^*).
\ee
These operators survive the quotient by null states to construct $\Hh_{BUFT}$, since if $|\omega]$ is a null state we have
\begin{align}\nonumber
[\omega|\hat{Z}(J)|J_1\ldots J_M]&=[\omega|JJ_1\ldots J_M]=0\\
[J_1\dots J_N|\hat{Z}(J)|\omega]&=[J^*J_1\dots J_N|\omega]=0.
\end{align}
Due to the bosonic nature of closed universe states we have\footnote{Note that this is a commutator, not an inner product!}
\be
[\hat{Z}(J),\hat{Z}(J')]=0,
\ee
which together with \eqref{conjZ} shows that the $\hat{Z}(J)$ are mutually commuting normal operators and thus can be simultaneously diagonalized (here we of course ignore any technical subtleties due to self-adjointness).  The eigenstates are conventionally called $|\alpha]$, with eigenvalues
\be
\hat{Z}(J)|\alpha]=Z_\alpha(J)|\alpha],
\ee
and the parameters which label them are called \textbf{$\alpha$-parameters}.  In principle there could be multiple $|\alpha]$ with the same eigenvalues $Z_\alpha(J)$, but in fact there are not since we have
\be\label{alphawf}
[J_1\ldots J_N|\alpha]=[HH|\hat{Z}(J_1)^\dagger\ldots \hat{Z}(J_N)^\dagger |\alpha]=Z_\alpha(J_1)^*\ldots Z_\alpha(J_N)^*[HH|\alpha]
\ee
so the eigenstates are determined by their eigenvalues. We will adopt a convention where the phase of $|\alpha]$ is chosen so that $[HH|\alpha]\geq 0$.

In what follows an important feature of $\alpha$-states is factorization: 
\be\label{alphafac}
[\alpha|\hat{Z}(J_1)\ldots \hat{Z}(J_N)|\alpha]=Z_\alpha(J_1)\ldots Z_\alpha(J_N).
\ee
In other words in an $\alpha$-state we can treat all spatial components of a closed universe as independent systems. In more general states this is not the case, and in particular in the (normalized) HH state we have
\be
\aleph^{-1}[HH|\hat{Z}(J_1)\ldots \hat{Z}(J_N)|HH]=\sum_\alpha p_\alpha Z_\alpha(J_1)\ldots Z_\alpha(J_N),
\ee
with
\be\label{palpha}
p_\alpha\equiv \frac{|[\alpha|HH]|^2}{\aleph}.
\ee
These $p_\alpha$ are positive and sum to one, so we can interpret this correlation function as arising from a classical average over factorizing states labeled by the $\alpha$-parameters.  A similar statement is true for general connected inner products:
\be\label{ipav}
\aleph^{-1}[J_1'\ldots J_N'|J_1\ldots J_M]=\sum_\alpha p_\alpha Z_\alpha(J_1')^*\ldots Z_\alpha(J_N')^*Z_\alpha(J_1)\ldots Z_\alpha(J_M).
\ee

\subsection{Averaged holography}\label{avsubsec}
Based on results analogous to \eqref{ipav}, in \cite{Coleman:1988cy} it was proposed that we should interpret the gravitational path integral as computing an average over fundamental quantum theories rather than the inner product in a fixed quantum theory.  In \cite{Coleman:1988cy} this was based on various approximations such as ignoring three-exit wormholes, but in \cite{Marolf:2020xie} it was shown using the argument given above that \eqref{ipav} holds exactly provided that the path integral is well-defined and positive-semidefinite. Moreover the factorization \eqref{alphafac} of expectation values of products of $\hat{Z}(J)$ in $\alpha$-states  can be viewed as evidence that we should view each $\alpha$-state as corresponding to some fixed holographic dual \cite{Marolf:2020xie}.\footnote{The observation of \cite{Saad:2019lba,Stanford:2019vob} that partition functions of JT gravity with any number of circle boundaries are computed by a random matrix theory averaging over Hamiltonians in some holographic dual can be viewed as a special case of this result, as can the relationship between the SYK model and the ``$G$'' and ``$\Sigma$'' variables \cite{Maldacena:2016hyu}.}  In terms of equations, the idea is that we should interpret \eqref{ipav} as telling us that
\be\label{BUFTah}
\aleph^{-1}[J_1'\ldots J_N'|J_1\ldots J_M]=\ol{\llangle J_1'\ldots J_N'|J_1\ldots J_M\rrangle},
\ee
where the double angle brackets indicate states in some fundamental holographic theory at fixed $\alpha$ and the bar indicates an average over $\alpha$.  
We will refer to this interpretation of the path integral as \textbf{averaged holography}, or AH for short. 

Although we motivated AH from BUFT, and they are sometimes conflated in the literature, there is an important difference between them that arises when we consider products of inner products. To make this explicit we first note that using the factorization equation \eqref{alphafac} we have
\be\label{alphafac2}
\llangle J_1'\ldots J_N'|J_1\ldots J_M\rrangle=Z_\alpha(J_1')^*\ldots Z_\alpha(J_N')^* Z_\alpha(J_1)\ldots Z_\alpha(J_M).
\ee
We therefore have
\be\label{AHvBUFT}
\ol{\llangle J_3|J_1\rrangle \llangle J_4|J_2\rrangle}=\ol{\llangle J_3J_4|J_1J_2\rrangle}=\aleph^{-1}[J_3J_4|J_1 J_2]\neq \aleph^{-2}[J_3|J_1][J_4|J_2],
\ee
where the first equality uses \eqref{alphafac2}, the second uses \eqref{BUFTah}, and the non-equality is the non-factorization of the right-hand side of \eqref{ipav}.  Thus if we wish to compute a product of inner products, we need to decide which one we mean: the product of inner products in BUFT (the right-hand side of \eqref{AHvBUFT}) or the average of the product of inner products in a fundamental holographic dual (the left-hand side of \eqref{AHvBUFT}).  In the path integral the difference arises because the latter includes topologies connecting boundaries in the different inner products while the former does not, see figure \ref{connectfig} for an illustration.  Another way to describe this situation is to say that the same gravitational path integral can be given inequivalent quantum interpretations: $\aleph^{-1}[J_3J_4|J_1J_2]$ is both an overlap of two-universe states in BUFT and the average of a product of one-universe overlaps in AH.  It is \textit{not} however a product of one-universe overlaps in BUFT.  A simple rule to remember is the following: in AH there is never more than one average over $\alpha$, while in BUFT there is an average for each inner product.
\bfig
\includegraphics[height=6cm]{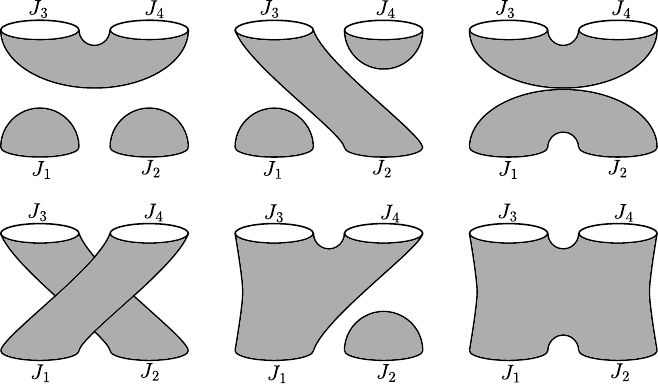}
\caption{Some of the topologies that contribute to $\aleph^{-1}[J_3J_4|J_1J_2]$ but not $\aleph^{-2}[J_3|J_1][J_4|J_3]$.}\label{connectfig}
\efig

There is a conceptual point which is worth mentioning here.  In the BUFT algorithm reviewed in the previous section, we start with a path integral and construct a Hilbert space.  For this algorithm to work however, we need the very nontrivial condition that the hermitian product \eqref{BUFTIP} is positive semidefinite.  Let's say instead that we start with averaged holography, so that the inner product by assumption has the form \eqref{BUFTah} with the factorization \eqref{alphafac2} for some function $Z_\alpha(J)$.  Then positive semidefiniteness is automatic: for any state
\be
|\psi]=\sum_N \int dJ_1\ldots dJ_N\psi_N(J_1,\ldots J_N)|J_1\ldots J_N]
\ee
we have
\begin{align}\nonumber
\aleph^{-1}[\psi|\psi]&=\sum_{N,N'}\int dJ_1\ldots dJ_NdJ_1'\ldots dJ_{N'}'\psi_{N'}(J_1,\ldots,J_{N'})^*\psi_N(J_1,\ldots J_N)\aleph^{-1}[J_1\ldots J_{N'}|J_1\ldots J_N]\\\nonumber
&=\sum_\alpha \Big|\sum_N \int dJ_1\ldots dJ_N\psi_N(J_1,\ldots J_N)Z_{\alpha}(J_1)\ldots Z_\alpha(J_N)\Big|^2\\
&\geq 0.
\end{align}
This can be viewed as evidence that AH is a more principled starting point than BUFT, we will return to this in section \ref{alphasec} below.

\section{Baby universe field theory vs averaged holography}
The difference between BUFT and AH emphasized in the last section has several applications to well-known problems in the literature.  In this section we discuss three of these.

\subsection{Closed universe Hilbert space}
Let's first review how the holographic one-dimensional Hilbert space claim arises in the context of BUFT and AH \cite{Marolf:2020xie}. From \eqref{alphafac2} together with \eqref{alphawf} we have
\be\label{ipfac}
\llangle J_1'\ldots J_N'|J_1\ldots J_M\rrangle=\frac{1}{p_\alpha \aleph}[J_1'\ldots J_N'|\alpha][\alpha|J_1\ldots J_M],
\ee
so the inner product between arbitrary states in fundamental Hilbert space factorizes.  In other words the inner product matrix has rank one, so when we quotient by null states we are left with a one-dimensional Hilbert space.  This formula gives the answer to question (1) from the introduction about the relationship between the ``one state'' of holography in a closed universe and the HH state: the state that the inner product factorizes on is $|\alpha]$, while $|HH]$ is a coherent superposition 
\be\label{HHsup}
|HH]=\sum_\alpha \sqrt{ p_\alpha\aleph}|\alpha]
\ee
of many $|\alpha]$ states.  

The details of $\alpha$-states are not needed to see that the inner product has rank one. Indeed if we introduce a condensed notation $\mathcal{J}$ for a set of closed universe sources with any number of spatial components, then for any states
\begin{align}\nonumber
|\psi\rrangle&=\int d\mathcal{J} \psi(\mathcal{J})|\mathcal{J}\rrangle\\
|\phi\rrangle&=\int d\mathcal{J} \phi(\mathcal{J})|\mathcal{J}\rrangle
\end{align}
we have
\be
\ol{\llangle \psi|\psi\rrangle \llangle \phi|\phi\rrangle-|\llangle \psi|\phi\rrangle|^2}=\aleph^{-1}[\psi\phi|\psi\phi]-\aleph^{-1}[\psi\phi|\phi\psi]=0
\ee
since the two terms describe path integrals with the same boundary conditions.  Moreover since
\be
\llangle \psi|\psi\rrangle \llangle \phi|\phi\rrangle-|\llangle \psi|\phi\rrangle|^2\geq 0
\ee
by the Cauchy-Schwartz inequality, this quantity must vanish without the average (or at least that the $\alpha$ where it doesn't vanish are measure zero).  Thus the transition probability between any two states in the fundamental Hilbert space is one, as one would expect from a theory with a one-dimensional Hilbert space.  

It is worth reflecting on why this argument does not work in CQG or BUFT.  After all closed universes are bosons also in CQG and BUFT, so we do have
\begin{align}\nonumber
\lan \psi\phi|\psi\phi\ran&=\lan \psi \phi|\phi\psi\ran\\
[\psi\phi|\psi\phi]&=[\psi \phi|\phi\psi].
\end{align}
What we do not have however is factorization, so these expressions do not tell us anything about $\lan \psi|\psi\ran \lan \phi|\phi\ran-|\lan \psi|\phi\ran|^2$ or $[\psi|\psi][\phi|\phi]-|[\psi|\phi]|^2$.  In order for these to vanish we would need to include topologies such as those shown in figure \ref{connectfig}, but in CQG and BUFT these are not included.  This is good of course, since BUFT has a large Hilbert space spanned by $\alpha$-states and the classical phase space of gravity in a closed universe is nontrivial (see for example \cite{Usatyuk:2024mzs,Alonso-Monsalve:2024oii} for the case of JT gravity with positive and negative cosmological constant).

We leave the physical interpretation of the one-dimensional Hilbert space in AH to section \ref{obsec} below.

\subsection{String worldsheet}
\bfig
\includegraphics[height=2.4cm]{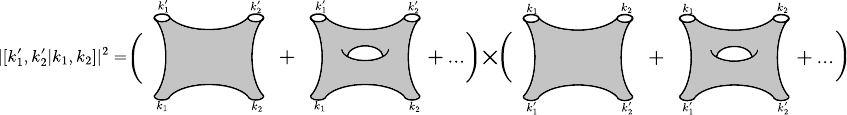}
\caption{Computing the square of a scattering amplitude in string theory using BUFT.  We emphasize that there are no worldsheets connecting the S-matrix and its complex conjugate.}\label{stringsfig}
\efig
In string theory the closed string scattering matrix is perturbatively computed by a sum over two-dimensional topologies with closed boundary insertions \cite{Polchinski:1998rq}.  In the present context we can think of this as a kind of closed universe quantum mechanics, and given our previous discussion we would like to know which one it is.  In the limit $g_s\to 0$ of vanishing string coupling only cylinders contribute (the one-boundary amplitude vanishes due to momentum conservation in the target space), so what we get is CQG.  What happens at finite $g_s$?  It clearly had better not be AH, as the Fock space of string scattering states is not one-dimensional.  So by process of elimination it must be BUFT, and indeed this is the case.  The key question is what we do when we compute the square of a string amplitude, for example in order to compute a differential cross section.  The answer of course is that we do \textit{not} include worldsheets that connect the S-matrix and its complex conjugate, see figure \ref{stringsfig} for an illustration, so we are indeed doing BUFT.  More properly, we are doing a version of BUFT where the boundaries are closed strings in the target space rather than Euclidean AdS boundaries.  In this context $\hat{Z}$ can be thought of as a closed string field that creates a loop of string in the target space, and the Hilbert space of BUFT is the Hilbert space of closed string field theory.  It would be interesting to further consider the target space interpretation of the $\alpha$ eigenstates of the closed string field, and in particular to understand the implications of factorization.  See \cite{Casali:2021ewu} for some discussion of the free worldline case.

Some readers may have found the previous paragraph confusing: isn't string theory supposed to be consistent with holography?  It is of course, but in string theory the spacetime that is holographic is the target space. String field theory is a perturbative formalism that cannot detect this; the statement here is only that there is no holography for the quantum gravity on the worldsheet. A related comment is the following: in \cite{Harlow:2025pvj} it was argued that the reason the Hilbert space of a holographic closed universe is one-dimensional is that there can be no ``super-observer'' who looks at the closed universe from the outside.  But in string theory there is one: the super-observer is us living in the target space looking at the strings!  So in this context it is quite natural that the super-observer can have a rich BUFT Hilbert space to do quantum mechanics on.  

\subsection{Evaporating black hole}\label{bhsec}
Recently there has been substantial progress on using the gravitational path integral to study the unitarity of black hole evaporation \cite{Penington:2019kki,Almheiri:2019qdq,Marolf:2020xie}.  These calculations however cannot really be understood without adapting a framework for the fundamental quantum origin of the gravitational path integral.  In higher dimensions this is complicated by the ambiguity of the path integral, which we will discuss further in section \ref{alphasec}, but even when the path integral is well-defined there is the question of whether we are doing CQG, BUFT, or AH.  This issue was discussed at length in \cite{Marolf:2020rpm}, which also pointed out the relevance of an earlier paper of Polchinski and Strominger \cite{Polchinski:1994zs} that presaged many of the recent developments.  In particular in the limit that the black hole is completely evaporated the ``replica wormhole'' calculations of \cite{Penington:2019kki,Almheiri:2019qdq}  coincide with the calculation of \cite{Polchinski:1994zs}, although they are not really under semiclassical control due to the Planckian singularity at the evaporation point.  In this subsection we will revisit  \cite{Polchinski:1994zs,Marolf:2020rpm} from the point of view of CQG, BUFT, and AH.  
\bfig
\includegraphics[height=6cm]{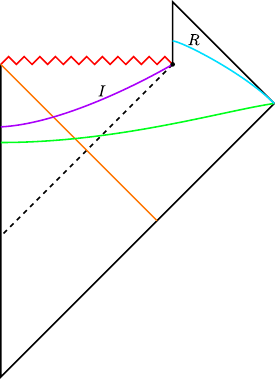}
\caption{The geometry of an evaporating black hole.  The information problem is usually studied on a ``nice'' Cauchy surface such as the green one shown here, but we will focus on a ``final'' Cauchy surface, consisting of the final radiation $R$ (shaded blue) together with a baby universe $I$ (shaded purple).}\label{bhevap1fig}
\efig
The geometry of an evaporating black hole is shown in figure \ref{bhevap1fig}.  Two Cauchy surfaces of interest are shown, a ``nice surface'' that avoids the Planckian region and a ``final'' Cauchy surface that does not.\footnote{Strictly speaking these are not Cauchy surfaces, since there are inextendible causal curves that begin or end at the evaporation point.  In a full quantum theory however whatever data is carried on those curves should radiate out into the evaporation region, so we will nonetheless call them Cauchy surfaces.}  The points we wish to make are most easily explained on the final Cauchy surface, which we will assume can be addressed using these path integral methods despite the singularity at the evaporation point.  As explained in \cite{Marolf:2020rpm} the ideas all generalize to the earlier Cauchy surface, at the cost of some inessential technical machinery.  The question we are of course interested in is whether the quantum state on the radiation $R$ part of the Cauchy surface is pure or mixed.  We can diagnose this by using the second Renyi entropy
\be
\Tr(\rho_R^2)=e^{-S_2(\rho_R)}.
\ee
\bfig
\includegraphics[height=8cm]{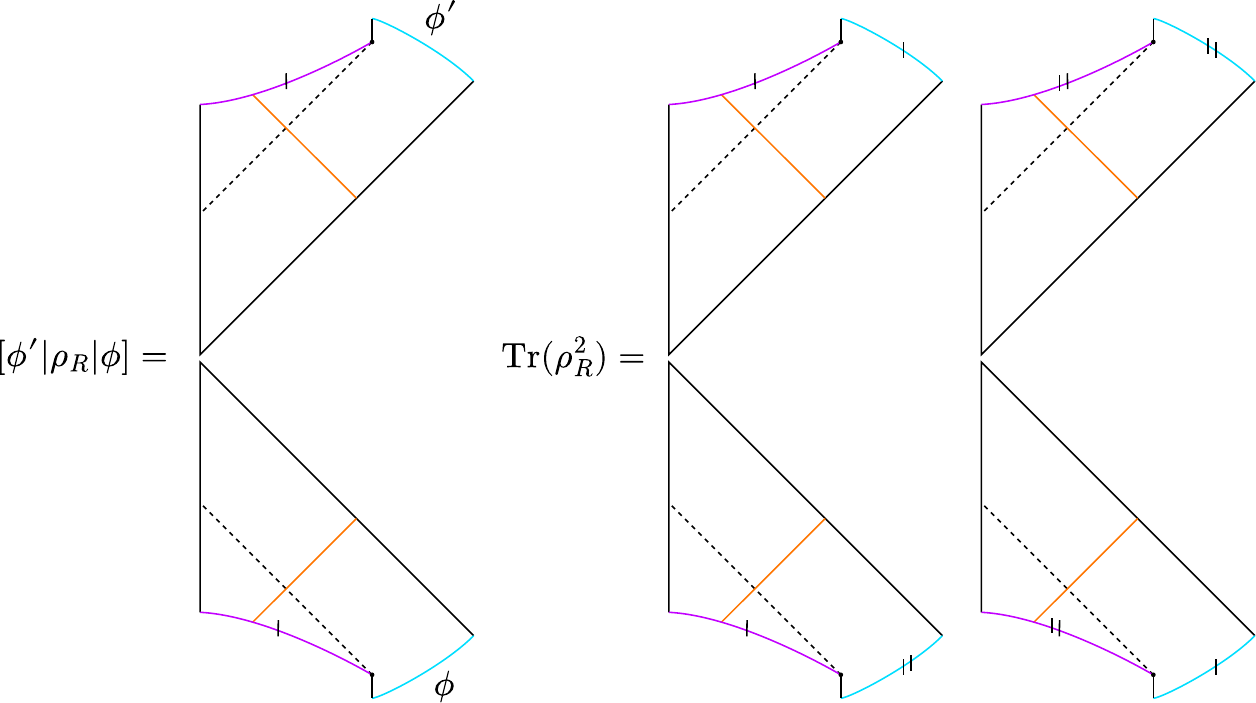}
\caption{Computing the state of the radiation for one completely evaporated black hole, and also its second Renyi entropy, in CQG or BUFT.  The lines crossing the blue and purple surfaces indicate identifications.}\label{bhevap2fig}
\efig
In CQG the radiation state and the second Renyi are computed by the path integral as in figure \ref{bhevap2fig}.  As expected they give Hawking's answer: the radiation is mixed with an entropy of order the coarse-grained entropy of the initial black hole,
\be\label{SCQG}
S_2(\rho_R)\sim S_{initial}.
\ee
This is also the answer in BUFT, since the calculation of $\rho_R$ as an operator on $\Hh_{BUFT}$ only involves a single-universe inner product.\footnote{Here we are assuming that there is no substantial disconnected contribution to the baby universe inner product for these states.  If there were then BUFT would already implement a version of the ``final state'' proposal of \cite{Horowitz:2003he}.  It would be interesting to see if there are situations where this possibility is realized.}  On the other hand if we compute the second Renyi using AH there is an additional contribution, shown in figure \ref{bhevap3fig}, and this gives 
\be
\ol{\Tr(\wt{\rho}_R^2)}\approx 1
\ee
and thus 
\be\label{SAH}
S_2(\wt{\rho}_R)\approx 0.
\ee
The reason that \eqref{SCQG} and \eqref{SAH} are different is precisely the difference between BUFT and AH: $\wt{\rho}_R$ is the radiation state in some fixed member of the ensemble, while $\rho_R$ is averaged over the ensemble and this introduces additional entropy.  In the path integral the difference arises because AH includes extra topologies that BUFT does not. 

\bfig
\includegraphics[height=7cm]{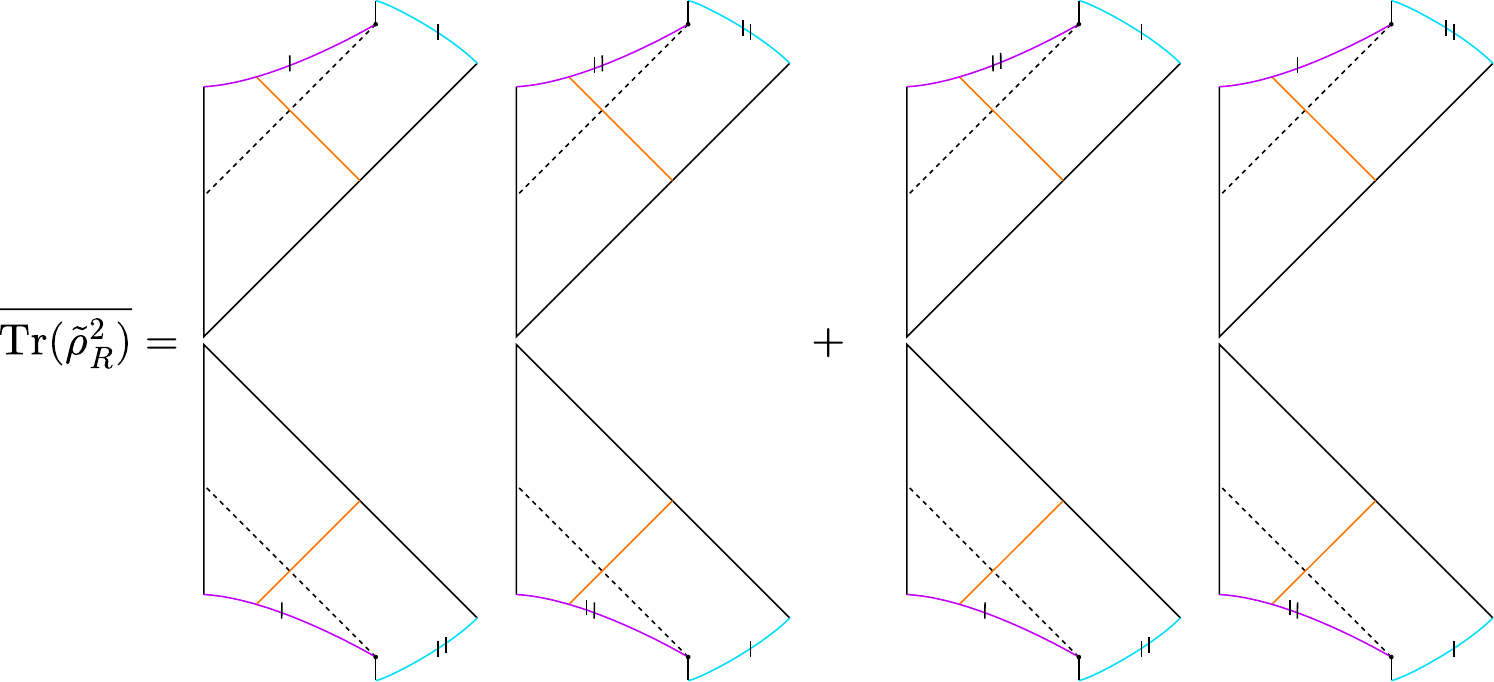}
\caption{Computing the second Renyi entropy in AH.  The new topology on the right switches which baby universe components are glued, leading to a contribution which is simply the norm of the Hawking state (which is of course normalized to one).}\label{bhevap3fig}
\efig
A source of potential confusion is that there \textit{is} a way to interpret the result \eqref{SAH} in BUFT even though BUFT predicts a mixed state for the radiation.  This is by way of the swap test, which uses the fact for any quantum state $\rho$ we have
\be\label{swap}
\Tr\left(\mathcal{S}(\rho\otimes \rho)\right)=\Tr(\rho^2)
\ee
where $\mathcal{S}$ is the unitary swap operator that exchanges the two copies of the system.  Thus $S_2(\rho)=0$ if and only if the expectation value of the swap operator is one.  In BUFT we can prepare a state with two identically-prepared evaporated black holes and compute the expectation value of $\mathcal{S}$, and the calculation is again given by figure \ref{bhevap3fig} so the result is close to one.  We thus get the ``pure state'' answer even though we know the state is mixed.  What happened?  The issue of course is the non-factorization of BUFT: equation \eqref{swap} only holds for states on the combined system that are a tensor product of identical one-system states, but in BUFT this is \textit{not} the case so we cannot use \eqref{swap}.\footnote{In \cite{Marolf:2020rpm} a ``swap entropy'' was introduced which is defined as minus the logarithm of the expectation value of $\mathcal{S}$ in a general quantum state on the doubled system.  We are not sure if this quantity is actually an entropy in any operational sense, but in any case the actual entropy of the radiation in BUFT is large.}  A BUFT state which \textit{is} compatible with the above data for $n$ evaporated black holes is \cite{Coleman:1988cy,Polchinski:1994zs,Marolf:2020rpm}
\be\label{psin}
|\psi]=\sum_\alpha \sqrt{p_\alpha}|\alpha]_I \otimes \left(|\psi_\alpha]_R\right)^{\otimes n},
\ee
where $p_\alpha$ is the initial distribution \eqref{palpha} for $\alpha$ in the HH state and $|\psi_\alpha]$ is the pure radiation state we would get for one black hole starting with some fixed $\alpha$.  The essential point here is that the baby universes left behind by the different black holes are \textit{not} independent systems, they all live in the same BUFT Hilbert space and thus are part of the same $|\alpha]$ state.  The  $|\psi_\alpha]$ are each normalized to one, but they are not necessarily orthonormal.  The state $|\psi]$ is indeed invariant under permutations of the $n$ black holes, which explains the swap expectation value being one for $n=2$, but this does NOT mean that tracing out the closed universe sector $I$ gives a pure state.   

It is interesting to consider the $n$-dependence of the Renyi entropy of the $n$ copies of the radiation in the state \eqref{psin}.  This is given by
\be\label{S2n}
S_2(\rho_{nR})=\sum_{\alpha,\beta} p_\alpha p_\beta\big|[\psi_\alpha|\psi_\beta]\big|^{2n}.
\ee
When $n=1$ we just get Hawking's answer \eqref{SCQG}.  If the different black holes interiors were really independent, we would expect the general answer to just be $n$ times the $n=1$ answer.  But that is not what we get: in the large $n$ limit the quantity $\big|[\psi_\alpha|\psi_\beta]\big|^{2n}$ becomes more and more sharply peaked at $\alpha=\beta$ (assuming that the $|\psi_\alpha]$ are at all distinguishable).  The final result depends on whether we assume the $\alpha$-parameters are discrete or continuous.  If they are discrete then the sum is just dominated by $\alpha=\beta$ and we get
\be
\lim_{n\to \infty}S_2(\rho_{nR})=\sum_\alpha p_\alpha^2,
\ee
which is just the second Renyi of the $\alpha$ distribution.\footnote{In the continouus case we instead can do the saddle point approximation to \eqref{S2n}, which leaves some residual logarithmic $n$-dependence, but the interpretation is similar \cite{Coleman:1988cy,Polchinski:1994zs}.}
In other words the uncertainty of the final state is completely controlled by our ignorance of the $\alpha$-parameters, rather than any non-unitarity at fixed $\alpha$, and if we measure the radiation of enough black holes to pin down the $\alpha$-parameters then we will collapse to an $\alpha$-state and the radiation will be pure for any additional black holes we create \cite{Polchinski:1994zs}.  In holography by contrast this process would be unnecessary, as we would be in a fixed ensemble realization from the start so the radiation state of the first black hole we evaporate would already be pure. We could confirm this purity by measuring the right rank-one projection on the radiation. This is why averaged holography gives a vanishing Renyi entropy \eqref{SAH} already for one black hole.  

The reader may at this point be wondering which is the option that we think is ``correct'': should we do CQG or BUFT or AH?  We will return to this question in section \ref{alphasec}, for now we are simply ``teaching the controversy''.  One thing that is worth emphasizing however is that we just saw that in BUFT the radiation state resulting from the evaporation of $n$ widely-separated and identically-prepared black holes is \textit{not} the tensor product of $n$ copies of the radiation state resulting from the evaporation of one black hole.  This is a violation of the principle of cluster decomposition, which says that the S-matrix should factorize for widely separated initial configurations, so one's credence in BUFT should be appropriately adjusted.\footnote{In perturbative string theory, where we argued that BUFT does need to be correct, cluster decomposition \textit{is} obeyed for the string S-matrix in the target space.  The cluster decomposition which fails is for some $1+1$ dimensional being living on the worldsheet.}

\section{A code interpretation of the closed universe path integral}\label{codesec}
So far we have been treating the gravitational path integral as the starting point for constructing a theory of quantum gravity.   Even in low-dimensional situations where this starting point is well-defined however, there is the unpleasant requirement that the BUFT hermitian product \eqref{BUFTIP} be positive semidefinite.  We saw at the end of section \ref{avsubsec} however that if we can find an alternative starting point that begins with an average over factorized holographic theories, then positive semidefiniteness is automatic.  In order to do this however we need some independent principle that computes for us the function $Z_\alpha(J)$, and we also need to select a probability distribution $p_\alpha$.  A nice way to organize this is using the idea of holographic error correcting codes \cite{Almheiri:2014lwa}.  From that point of view the goal is to construct a linear holographic encoding map
\be
V_\alpha:\mathcal{H}_{BUFT}\to \mathcal{H}_{fund},
\ee
that acts as
\be
V_\alpha|\psi]=|\psi\rrangle.
\ee 
We emphasize that this holographic encoding is happening with some particular choice of $\alpha$.  Due to factorization in the fundamental description, $V_\alpha$ is determined by its action on one-universe states:
\be
V_\alpha|J_1\ldots J_N]=|J_1\ldots J_N\rrangle=|J_1\rrangle \ldots |J_N\rrangle=V_\alpha|J_1\rrangle \ldots V_\alpha|J_N\rrangle.
\ee
In this section we will discuss some simple models where we can give a direct proposal for $V_\alpha$ and then see what bulk theory emerges from it.  
Our criterion for success in this section will be showing that
\be
\sum_\alpha p_\alpha [\phi|V_\alpha^\dagger V_\alpha|\psi]\approx\frac{1}{\aleph}[\phi|\psi].
\ee
This is not the usual holographic encoding condition, which would not include the average over $\alpha$ on the left-hand side, but it is the thing that most closely matches the gravitational path integral in situations where it is well-defined.  In section \ref{obsec} below we will discuss what happens when we do not average over $\alpha$.

Before beginning with the models, it is worth mentioning that in situations where $V_\alpha$ does arise from a positive semi-definite path integral, from \eqref{ipfac} we simply have
\be\label{Va}
V_\alpha=\frac{1}{\sqrt{p_\alpha \aleph}}[\alpha|.
\ee
This clearly shows the one-dimensional nature of $\mathcal{H}_{fund}$.  In particular acting on a one-universe state we have
\be
V_\alpha|J]=Z_\alpha(J),
\ee
so $V_\alpha$ and $Z_\alpha(J)$ contain the same information.  Moreover from \eqref{Va} and \eqref{palpha} we have
\be\label{HH1}
|HH\rrangle=V_\alpha|HH]=1,
\ee
so in the fundamental description the Hartle-Hawking state is just one! It is also interesting to note that
\be
|\beta\rrangle=V_\alpha|\beta]=\frac{\delta_{\alpha\beta}}{\sqrt{p_\alpha \aleph}},
\ee
so if we try to map the ``wrong'' $|\alpha]$ state to the fundamental description we get zero.

\subsection{One-particle topological model}
\bfig
\includegraphics[height=4.5cm]{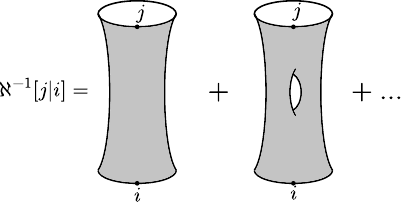}
\caption{Computing the BUFT inner product in the one-particle topological model.}\label{1pipfig}
\efig
Perhaps the simplest model where BUFT/AH can be studied to the end is the topological model of \cite{Marolf:2020xie}, where we sum over $1+1$ dimensional geometries $\M$ with a Euclidean action given by
\be\label{topmod}
S_E(\M)=-S_0\chi(\M).
\ee
Here $\chi(\M)$ is the Euler character of $\M$.  This model however has no local degrees of freedom so there is nothing to discuss about the physics in a single closed universe (the only degree of freedom is the number of spatial components).  In \cite{Usatyuk:2024mzs} a generalization of this model was introduced to fix this, consisting of the topological model together with a matter wordline carrying a species index $i$.  For example in this model the BUFT transition amplitude between two one-universe states, one with a particle in state $i$ and the other with a particle in state $j$, is
\be\label{1pip}
\aleph^{-1}[j|i]=\frac{\delta_{ij}}{1-e^{-2S_0}},
\ee
with the numerator being the worldline species propagator $\delta_{ij}$ and the denominator giving the sum over genus.  See figure \ref{1pipfig} for an illustration.

\bfig
\includegraphics[height=5cm]{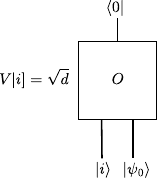}
\caption{A holographic map for encoding a single-particle one-universe state.}\label{1pcodefig}
\efig
In \cite{Harlow:2025pvj} this topological model with worldline matter was given an averaged holography interpretation by introducing a holographic encoding map that acts on one-particle single-universe states as
\be
|i\rrangle\equiv V_O|i]\equiv \sqrt{d}\lan 0|O|i\ran|\psi_0\ran,
\ee
where $O$ is a $d\times d$ orthogonal matrix chosen at random in the Haar measure and $|\psi_0\ran$ is some fixed input state into additional input legs of $O$ so that we can take the large $d$ limit. See figure \ref{1pcodefig} for an illustration.  We emphasize that $|i\rrangle$ is a number, so the fundamental Hilbert space is indeed one-dimensional.   The inner product \eqref{1pip} is then recovered up to exponentially small corrections (that we will not try to capture) by standard Haar integration technology after averaging over $O$:
\be
\ol{\llangle j|i\rrangle}=\int dO [j|V_O^\dagger V_O|i]=\lan j|i\ran=\delta_{ij} \approx \aleph^{-1}[j|i].  
\ee
Comparing to equation \eqref{ipav}, we see that in this model $O$ indeed plays the role of the $\alpha$-parameters and 
\be
Z_O(i)=V_O|i]=|i\rrangle.
\ee
At large $d$ this agreement continues for multi-universe amplitudes, for example we have
\be
\ol{\llangle i'j'|ij\rrangle}=\int dO[i'j'|V_O^\dagger V_O|ij]=\delta_{ii'}\delta_{jj'}+\delta_{ij'}\delta_{ji'}+\delta_{ii'}\delta_{jj'}+O(1/d^2),
\ee
which matches the bulk calculation as shown in figure \ref{1p2ipfig}.  See appendix A of \cite{Harlow:2025pvj} for the rules for doing these Haar integrals.  Here the encoding map of a two-universe state is just two copies of the one-universe encoding:
\be
V_O|ij]=|ij\rrangle=|i\rrangle|j\rrangle=\sqrt{d}\lan 0|O|i\ran|\psi_0\ran\times \sqrt{d}\lan 0|O|j\ran|\psi_0\ran,
\ee
as required by factorization.  We emphasize that in the definition of $V_O$ we have used the CQG state $|i\ran$ rather than the BUFT state $|i]$, the left line going into $O$ is a CQG Hilbert space line. 
\bfig
\includegraphics[height=3cm]{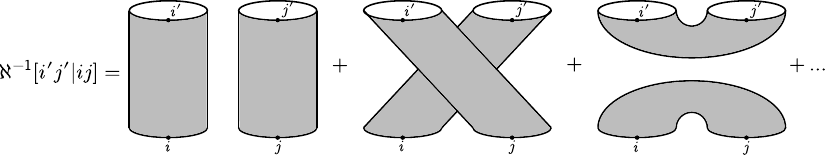}
\caption{Computing a two-universe inner product in the topological model with a worldline particle.}\label{1p2ipfig}
\efig

For our current purposes however this model has an important defect: in the bulk calculation of the inner product the disconnected contribution arising from two disks vanishes due to the worldline propagator having nowhere to go.  This leads to the one-particle state having zero overlap with the Hartle-Hawking state:
\be
[i|HH]=0.  
\ee
In the code model this equation follows because an integral over a single Haar random $O$ vanishes by the symmetry $O\to -O$ of the Haar measure.  To learn about the HH state and disconnected contributions to the BUFT inner product, which is one of the main goals of this paper, we need a better model.  

\subsection{Two-particle topological model}
\bfig
\includegraphics[height=7.5cm]{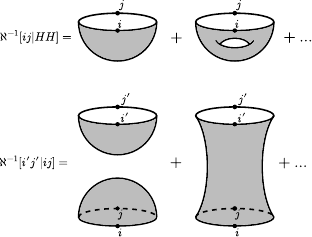}
\caption{Computing the BUFT HH state and one-universe inner product for two-particle states.}\label{2pipfig}
\efig
The simplest thing we can do to get a nonzero HH overlap and a nonvanishing disconnected contribution to the one-universe inner product is consider two-particle states instead of one-particle states.  In the bulk we then have\footnote{This maximally-entangled state is reminiscent of the infinite-temperature state of the de Sitter static patch argued to exist in \cite{Chandrasekaran:2022cip}.  We are not sure if this analogy between our two particles and two entangled static patches is a good one however.}
\be\label{HHtop}
\aleph^{-1}[ij|HH]=\frac{e^{S_0}}{1-e^{-2S_0}}\delta_{ij} 
\ee 
and 
\be\label{2btop}
\aleph^{-1}[i'j'|ij]=\left(\frac{e^{S_0}}{1-e^{-2S_0}}\right)^2\delta_{ij}\delta_{i'j'}+\frac{1}{1-e^{-2S_0}}\left(\delta_{ii'}\delta_{jj'}+\delta_{ij'}\delta_{ji'}+\delta_{ij}\delta_{i'j'}\right),
\ee
where the denominator factors arise from summing over genus.  We emphasize that now $|ij]$ is a two-particle state in one universe instead of two-universe state each with one particle.  See figure \ref{2pipfig} for an illustration.  The exponentially enhanced term in the inner product comes from the disconnected contribution.  The proposed encoding map for this theory is shown in figure \ref{2pcodefig}: each two-particle one-universe state is mapped to a number via
\be
|ij\rrangle=V_O|ij]=d\lan \chi|\left(O\otimes O\right)\left(|i\ran|\psi_0\ran \otimes |j\ran|\psi_0\ran\right),
\ee
where $O$ is again a Haar-random $d\times d$ orthogonal matrix and $|\chi\ran$ is an entangled state that is invariant under the swap operation exchanging the two $d$-dimensional tensor factors.  We will work in the limit $d\to \infty$, keeping the range of the flavor index and also the overlap
\be
\lan \chi|\mathrm{max}\ran\equiv \sqrt{k}
\ee
both finite.  Here $|\mathrm{max}\ran$ is the \textit{unnormalized} maximally mixed state 
\be
|\mathrm{max}\ran=\sum_{n=1,d}|n\ran\otimes |n\ran.
\ee
As before the encoding of multi-universe states factorizes:
\be
V_O|ij,i'j']=|ij,i'j'\rrangle=|ij\rrangle |i'j'\rrangle.
\ee

\bfig
\includegraphics[height=6cm]{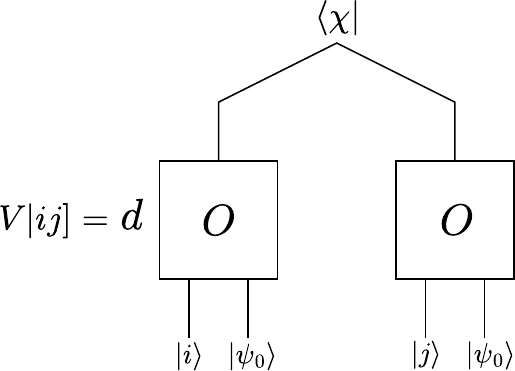}
\caption{A holographic map for encoding a two-particle one-universe state.}\label{2pcodefig}
\efig
As a first calculation in this model we can compute the overlap of the HH state with a one-boundary state:
\be\label{2pHH}
\ol{\llangle ij|HH\rrangle}=\int dO [ij|V_O^\dagger=\lan \mathrm{max}|\chi\ran\delta_{ij}=\sqrt{k}\delta_{ij}.
\ee
Here we have used \eqref{HH1}.  Thus to match \eqref{HHtop} up to subleading terms in $e^{-2S_0}$, we should take
\be
k=e^{2S_0}.
\ee
The two-boundary connected amplitude is given by
\be
\ol{\llangle i'j'|ij\rrangle}=\int dO [ i'j'|V_O^\dagger V_O|ij]=k\delta_{ij}\delta_{i'j'}+\delta_{ii'}\delta_{jj'}+\delta_{ij'}\delta_{ji'},
\ee
which agrees with \eqref{2btop} at leading order in $1/k$ for each index structure.

This agreement continues for connected amplitudes with more boundaries provided that we give $|\chi\ran$ a little more structure.  Namely we take it to be maximally entangled on a pair of subspaces of rank $k$, so that index contractions involving $|\chi\ran$ become a simple matter of counting loops and supplying a factor of $k$ for each loop and $1/\sqrt{k}$ for each $|\chi\ran$.  For example we have the three-boundary connected amplitude:
\begin{align}\nonumber
\ol{\llangle HH|i_1j_1,i_2j_2,i_3j_3\rrangle}&=\int dO \, V_O |i_1j_1,i_2j_2,i_3j_3]\\\nonumber
=&k^{3/2}\delta_{i_1j_1}\delta_{i_2j_2}\delta_{i_3j_3}+k^{1/2}\left(\delta_{i_1j_1}\left(\delta_{i_2i_3}\delta_{j_2 j_3}+\delta_{i_2j_3}\delta_{j_2 i_3}\right)+1\to 2\to 3+1\to 3\to 2\right)\\
&+k^{-1/2}\left(\delta_{i_1i_2}\left(\delta_{j_2 i_3}\delta_{j_3 j_1}+\delta_{j_2 j_3}\delta_{i_3 j_1}\right)+\delta_{i_1j_2}\left(\delta_{i_2 i_3}\delta_{j_3 j_1}+\delta_{i_2 j_3}\delta_{i_3 j_1}\right)+2\leftrightarrow 3\right),
\end{align}
where the first line matches the four disconnected contributions to the path integral and the second line matches the fully connected ``pair of pants'' geometry, again with the match working at leading order in $1/k$ for each index structure.

\section{Do $\alpha$-parameters exist?}\label{alphasec}
It is now time for us to confront the fact that outside of low-dimensional models the gravitational path integral is not well-defined.  To start with it is not renormalizable, and so must be viewed as an effective theory with unknown higher-derivative corrections.  To make things worse, the sum over topologies is quite unlikely to converge (already in $1+1$ dimensions the string perturbation series is divergent due to the factorial growth of moduli space volumes \cite{Shenker:1990uf}). We thus find ourselves faced with an unpleasant question: how much of the story we just told about CQG and BUFT and AH can actually be trusted outside of soluble models in low spacetime dimension?\footnote{The situation was nicely summarized by Coleman \cite{Coleman:1988tj}: ``The Euclidean formulation of gravity is not a subject with firm foundations and clear rules of procedure; indeed, it is more like a trackless swamp.  I think I have threaded my way through it safely, but it is always possible that unknown to myself I am up to my neck in quicksand and sinking fast.''}  One might hope that the answer to this question is ``most of it'', for example that seems to be the point of view of \cite{Marolf:2020xie,Marolf:2020rpm}, but in our view the evidence points strongly in the other direction.  

\subsection{The trouble with $\alpha$}
As one clear sign of trouble we can consider the pure topological model of \cite{Marolf:2020xie} with action \eqref{topmod}.  If we study this model in the approximation where only bipartite wormholes are considered, as in \cite{Coleman:1988cy}, then the spectrum of the $\hat{Z}$ operator is continuous and runs from $-\infty$ to $\infty$ (it is essentially the $X$ operator for a harmonic oscillator).  One might naively think that including the other topologies which this approximation neglects should only result in small corrections, since they are exponentially suppressed in $S_0$, but in fact the full solution of the model tells us that the spectrum of $\hat{Z}$ is discrete and bounded from below \cite{Marolf:2020xie}.  Apparently the structure of the $|\alpha]$ states depends very sensitively on the full set of topologies, and in particular on the sum being convergent.  Indeed in a setting where the path integral is not well-defined, our view is that there is no reason to expect the existence of $|\alpha]$ states at all.

In addition to attacking the argument for $\alpha$-parameters, we can also complain about their consequences.  Indeed as we reviewed in section \ref{bhsec}, including a sum over $\alpha$-parameters leads to a mixed state of the radiation from an evaporating black hole, as well as a violation of cluster decomposition in the black hole S-matrix.  

The case against $\alpha$-parameters becomes even more damning if we take into account evidence from string theory and holography \cite{McNamara:2020uza}.  Indeed string theory as far as we know is a theory with no dimensionless parameters, since every low-energy coupling constant is the expectation value of a field.  For example the string coupling constant is the expectation value of the dilaton field \cite{Polchinski:1998rq}. Similarly in AdS/CFT any parameter of the dual CFT is a boundary condition for some field in the bulk.   For example in the $\mathcal{N}=4$ super Yang Mills theory dual to IIB string theory in $AdS_5\times \mathbb{S}^5$,  the gauge coupling $g$ is the square root of the boundary value of the dilaton and the number $N_c$ of colors is the boundary condition for the five-form flux wrapping the $\mathbb{S}^5$ \cite{Maldacena:1997re}.  More generally, if we have two different CFTs then on a spatially disconnected manifold we can always have one CFT on one component and the other CFT on another component, which must be dual to some bulk configuration that interpolates between them.  For example if we have a pair of boundaries with different values of $N_c$, then there is a bulk configuration with Euclidean D3 branes that connects them.  

We can use an argument of \cite{Usatyuk:2024mzs} to make the inconsistency of $\alpha$-parameters and AdS/CFT even more explicit.  Returning to the BUFT inner product between one-universe states prepared by Euclidean AdS boundaries, the path integral rule is to sum over all Euclidean geometries connecting the two boundaries.  AdS/CFT tells how to do this calculation non-perturbatively in IIB string theory, the answer is:
\be
[J'|J]=Z(J')^* Z(J),
\ee
where $Z(J)$ is the Euclidean partition function of the CFT as a function of sources $J$.  This agrees with \eqref{ipav} \textit{only} if there is no sum over $\alpha$-parameters.  Indeed we can think of AdS/CFT as giving us precisely the needed input for the alternative direction of the logic advocated at the beginning of section \ref{codesec}: rather than deriving $Z_\alpha(J)$ and $p_\alpha$  from the path integral, we derive the path integral from $Z_\alpha(J)$ and $p_\alpha$.  But it has given them to us with a twist: there is only one $\alpha$, so $p_\alpha=1$ and $Z_\alpha(J)=Z(J)$. 

\subsection{Averaging is for sources}
Given the previous subsection, you may be wondering why we spent such a long time discussing BUFT and $\alpha$-parameters.  One reason is that there is at least one case, string perturbation theory, where BUFT really is the right formalism.  Another reason is that one hears a lot of discussion about $\alpha$-parameters these days, so it is good to be clear about how they work and we hope we have demystified some confusing features in our exposition.  For our purposes however the main reason is different: although a model with many $\alpha$-parameters is probably a bad model for how quantum gravity works in more realistic settings, the behavior of such a model in an $|\alpha]$ state seems to be a \textit{good} model for higher-dimensional holography.  Indeed in an $|\alpha]$ state we have factorization and also a unitary black hole S-matrix that obeys cluster decomposition.  The central question before us therefore is to what extent ordinary semiclassical physics can emerge in a holographic theory with no $\alpha$-parameters.  Since a naive analysis of the path integral led to many $\alpha$-parameters, it is not at all clear that this emergence can succeed.  In fact Hawking's information problem can be ahistorically thought of as illustrating the tension between holography and $\alpha$-parameters.  We can study this question using our above models in an $|\alpha]$-state, which is why they are not useless.

In simple AdS calculations such as correlation functions of light fields in the vacuum, the contributions of higher topologies are tiny, of order $e^{-N_c^2}$, so the state of the closed universe sector does not contribute in any substantial way.  We thus should expect concordance between (unaveraged) holography, CQG, and BUFT, and indeed that is what is found.  A more nontrivial calculation is that of the black hole partition function \cite{Gibbons:1976ue,Hawking:1982dh}, which goes beyond CQG, but there is still only a single boundary component so there is no tension with factorization and BUFT and holography are in agreement.  

In order to get a clear factorization problem we should consider a quantity with multiple AdS boundaries.  Perhaps the simplest such quantity where we can indeed find a problem is the \textbf{spectral form factor} \cite{Papadodimas:2015xma,Cotler:2016fpe}, which in CFT language is defined by
\be
g(t)=|\Tr_{\partial\Sigma}\left(e^{-(\beta+it)H_{\partial\Sigma}}\right)|^2,
\ee
where the trace is computed in the CFT Hilbert space with spatial manifold $\partial \Sigma$ and $H_{\partial \Sigma}$ is the CFT Hamiltonian on $\partial \Sigma$.  By construction the spectral form factor is the square of a CFT partition function:
\be
g(t)=Z(\beta+it,\partial \Sigma)Z(\beta-it,\partial \Sigma),
\ee
where $Z(\beta,\partial \Sigma)$ is the Euclidean CFT partition function on $\mathbb{S}^1\times \partial \Sigma$ with circle radius $\beta$.  In particular when $t=0$ this quantity is the same as what we called the norm of a one-universe state in the closed universe inner product:
\be
g(0)=\llangle\beta,\partial\Sigma|\beta,\partial\Sigma\rrangle.
\ee
More generally we can think of the spectral form factor as an analytically continued version of this inner product to complex spatial geometry:
\be
g(t)=\llangle\beta-it,\partial \Sigma|\beta+it,\partial \Sigma\rrangle.
\ee
The question of interest here is how $g(t)$ compares to the gravitational path integral evaluated with these boundary conditions, which we can write as $\aleph^{-1}[\beta-it,\partial\Sigma|\beta+it,\partial \Sigma]$.  When the dominant contribution to this path integral is disconnected, then the connected contributions only lead to small violations of factorization that arguably are within the allowed margin of error for effective field theory.  For example in the two-dimensional topological model we simply have (see \eqref{2btop} but without the two matter particles)
\be
g(t)=\left(\frac{e^{S_0}}{1-e^{-2S_0}}\right)^2+\frac{1}{1-e^{-2S_0}},
\ee
where the first term comes from two disks and the second from a cylinder.  The second term is a violation of factorization, but it is exponentially small compared to the first term.  Given that the bulk is at best emergent anyways, perhaps we do not mind.  To really get into trouble therefore, we need find some choice of boundary sources $J$ where the disconnected contribution is small.  In \cite{Cotler:2016fpe} and then \cite{Saad:2018bqo,Saad:2019lba} it was beautifully explained that this indeed happens in the limit of large $t$ in the spectral form factor.  For example in JT gravity the disk and cylinder partition functions are \cite{Saad:2019lba}
\begin{align}\nonumber
Z_{disk}(\beta)&=\sqrt{\frac{4\phi_b^3}{\pi \beta^3}}e^{S_0+4\pi^2\phi_b/\beta}\\
Z_{cylinder}(\beta_1,\beta_2)&=\frac{\sqrt{\beta_1\beta_2}}{2\pi(\beta_1+\beta_2)},
\end{align}
where $\phi_b$ is a constant appearing in the dilaton boundary condition $\phi_{boundary}=\phi_b/\epsilon$, so we have
\be\label{bulksff}
g(t)\approx \frac{4\phi_b^3}{\pi (\beta^2+t^2)^{3/2}}e^{2S_0+\frac{8\pi^2\phi_b}{\beta^2+t^2}}+\frac{\sqrt{\beta^2+t^2}}{4\pi \beta}.
\ee
Thus at large $t$ the disk contribution decreases and the cylinder contribution increases, so when $t\gtrsim e^{S_0/2}$ the cylinder indeed becomes dominant and non-factorization is unavoidable.\footnote{Of course at this point other saddles also become important, so to get a sharp tension we should approach this time but not quite get there.}  What happens in the holographic theory was studied numerically in \cite{Cotler:2016fpe}, essentially as we get to times of order $e^{S_0/2}$ then $g(t)$ begins to undergo large erratic fluctuations as a function of time.  In particular it does \textit{not} agree with the simple bulk answer \eqref{bulksff}.  This is as it must be, since the boundary calculation manifestly factorizes.  In situations where there is an ensemble of boundary theories this can be fixed by averaging over $\alpha$ parameters, but fortunately this is not the only option: we can instead just average $g(t)$ over an appropriate time window.  This is the key lesson: in a fixed holographic theory in higher dimensions we cannot average over $\alpha$ since it does not exist, but we \textit{can} average over $J$ and this can be enough to restore agreement with the path integral.

We do need to be careful to acknowledge however that averaging over $J$ is a physical operation, and we should only do it if the question we are interested in calls for it.  In this our perspective diverges somewhat from that of \cite{Liu:2025cml,Liu:2025ikq} (see also \cite{Kudler-Flam:2025cki}), where it was proposed that some kind of $J$ averaging, called a ``filter'' in \cite{Liu:2025ikq}, should be part of the fundamental holographic dictionary.  Our view is instead that the dictionary is the usual one, but that some quantities of interest are naturally defined to include some kind of average over $J$. Sometimes this averaging is enough to recover the path integral and sometimes it isn't, and when it is recovered sometimes it is only to within some accuracy that is controlled by the amount of averaging (we will see this explicitly in the next subsection).  


\subsection{The observer rule as averaging over $J$}
In the context of closed universe cosmology it is even easier to get into situations where the disconnected contribution to the path integral is small compared to the connected one.  For example in the one-particle cosmological model of section \ref{codesec} the disconnected contribution to the inner product automatically vanishes, and in  two-particle cosmological model the same is true in \eqref{2btop} provided that we have $i\neq i'$ or $j\neq j'$.  This problem was discussed in \cite{Harlow:2025pvj}, where it was proposed to restore the gravitational path integral interpretation by introducing an observer in the closed universe and then including in the inner product a decohering channel in their pointer basis.  We now argue that this decohering channel implements an averaging over $J$ which is quite similar to what we just discussed for the spectral form factor, and that it serves basically the same purpose.

\bfig
\includegraphics[height=3.5cm]{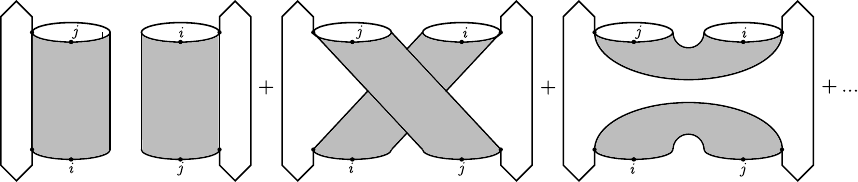}
\caption{Applying the observer rule to the squared inner product  of one-particle one-universe states.  The first configuration is enhanced relative to the second two by a factor of $e^{S_{Ob}}$ since it has two observer index loops while the second two only have one.  This suppresses the variance in $|\llangle j|i\rrangle|^2$, leading to answers at fixed $O$ that are consistent with the path integral.}\label{1p2ipobfig}
\efig
In the one-particle topological model we can illustrate this quite precisely by re-purposing the calculation shown in figure \ref{1p2ipfig} to tell us that
\be\label{largevar1}
\ol{\big|\llangle j|i\rrangle-\ol{\llangle j|i\rrangle}\big|^2 }=1+\delta_{ij}+O(e^{-2S_0}),
\ee
so the deviation of the inner product $\llangle j|i\rrangle$ from its bulk value of $\delta_{ij}$ is $O(1)$ for a typical choice of $O$. This is the same phenomenon as the large erratic fluctuations in the late-time spectral form factor.  What the observer decoherence rule of \cite{Harlow:2025pvj} does in practice is ``decorate'' the path integral calculation by feeding in an entangled state of the observer $Ob$ and a ``clone'' $Ob'$ in some external system, see figure \ref{1p2ipobfig} for an illustration of how this modifies the calculation of $\ol{|\llangle j|i\rrangle|^2}$.  The result of this decoration is to suppress the fluctuations about the bulk value exponentially in the observer entropy:
\be\label{obvar1}
\ol{\big|\llangle j|i\rrangle-\ol{\llangle j|i\rrangle}\big|^2 }=e^{-S_{Ob}}\left(1+\delta_{ij}\right)+O(e^{-2S_0}).
\ee
In \cite{Harlow:2025pvj} it was argued that this suppression is sufficient to account for a valid semiclassical experience for this observer, since anyways they cannot do physics to a precision better than $e^{-S_{Ob}}$.  

The main point we want to make here is that if we trace out the cloned system $Ob'$, we can view the observer rule as implementing an average over the internal state of the observer in their pointer basis.  Since the observer is created in the past and annihilated at the future by CFT sources, with the source depending on the internal state of the observer, we can view this averaging as a special type of averaging over $J$.  We emphasize that we have not changed the rules for evaluating the path integral or the CFT partition function, we are merely averaging over boundary conditions / CFT sources.  We would like to contrast this with the alternative observer rule proposed in \cite{Abdalla:2025gzn}, which instead proposes just dropping the second two contributions in figure \ref{1p2ipobfig} from the gravitational path integral.  That rule would follow from the rule of \cite{Harlow:2025pvj} in the limit $S_{Ob}\to \infty$, but it is not so clear that this limit can be taken since it would likely require the observer to be infinitely heavy.  We are not currently aware of a CFT prescription (such as some other $J$ averaging) that would lead to the rule of  \cite{Abdalla:2025gzn}.

\section{Closed universe inner product in a fixed holographic theory}\label{obsec}
In this section we will revisit the topological two-particle model of section \ref{codesec}, now from the perspective of understanding to what extent BUFT and/or CQG emerge when the encoding orthogonal matrix $O$ is fixed rather than averaged over.  In other words, we will study to what extent we have
\be\label{hbuqm}
\llangle \phi|\psi\rrangle \approx \aleph^{-1}[\phi|\psi]
\ee
or
\be\label{hcqg}
\llangle \phi|\psi\rrangle \approx \lan \phi|\psi\ran.
\ee
Our conclusion will be that adopting the observer rule the former holds up to errors which are of order $e^{-S_{Ob}/2}$, while the latter only holds if the cylinder dominates the one-universe inner product in BUFT.  

\subsection{Computing the variance without an observer}

We'll begin by computing the variance of $\llangle ij|HH\rrangle$.  Since we already showed that the path integral model emerges from the code model we can just use the former, which gives
\be
\ol{\Big|\llangle ij|HH\rrangle-\ol{\llangle ij|HH\rrangle}\Big|^2}=\aleph^{-1}[ij|ij]-\aleph^{-2}|[ij|HH]|^2=1+2\delta_{ij}.
\ee
In the last step we used \eqref{HHtop} and \eqref{2btop} and neglected terms that are of order $e^{-2S_0}$.  This variance is of order one, but we should remember that we are comparing it to 
\be\label{HH6}
\ol{\llangle ij|HH\rrangle}=\frac{e^{S_0}}{1-e^{-2S_0}}\delta_{ij},
\ee
so the variance is exponentially small compared to the signal when $i=j$.  Thus for a typical fixed $O$ we have
\be
\llangle ii|HH\rrangle\approx \aleph^{-1}[ii|HH].
\ee
On the other hand when $i\neq j$ then $[ij|HH]=0$ and the variance is large.  These results are similar to the early and late time situations for the spectral form factor.

This is a good point to stop and revisit question (1) from the introduction: how does the ``one state'' $V_O^\dagger$ in the holographic encoding map relate to the HH state?  We are now working in a fixed holographic theory, so there is no large set of of $|\alpha]$-states to expand $|HH]$ in.  Nonetheless the state $V_O^\dagger$ is well-defined, and it is not equal to the HH state.  Indeed in this model the one-universe part of the latter is simply given by the maximally entangled state \eqref{HH6}, independent of the details of $O$.  The HH state is instead a kind of low-energy coarse-grained version of the unique state, but they are not the same.

\bfig
\includegraphics[height=11cm]{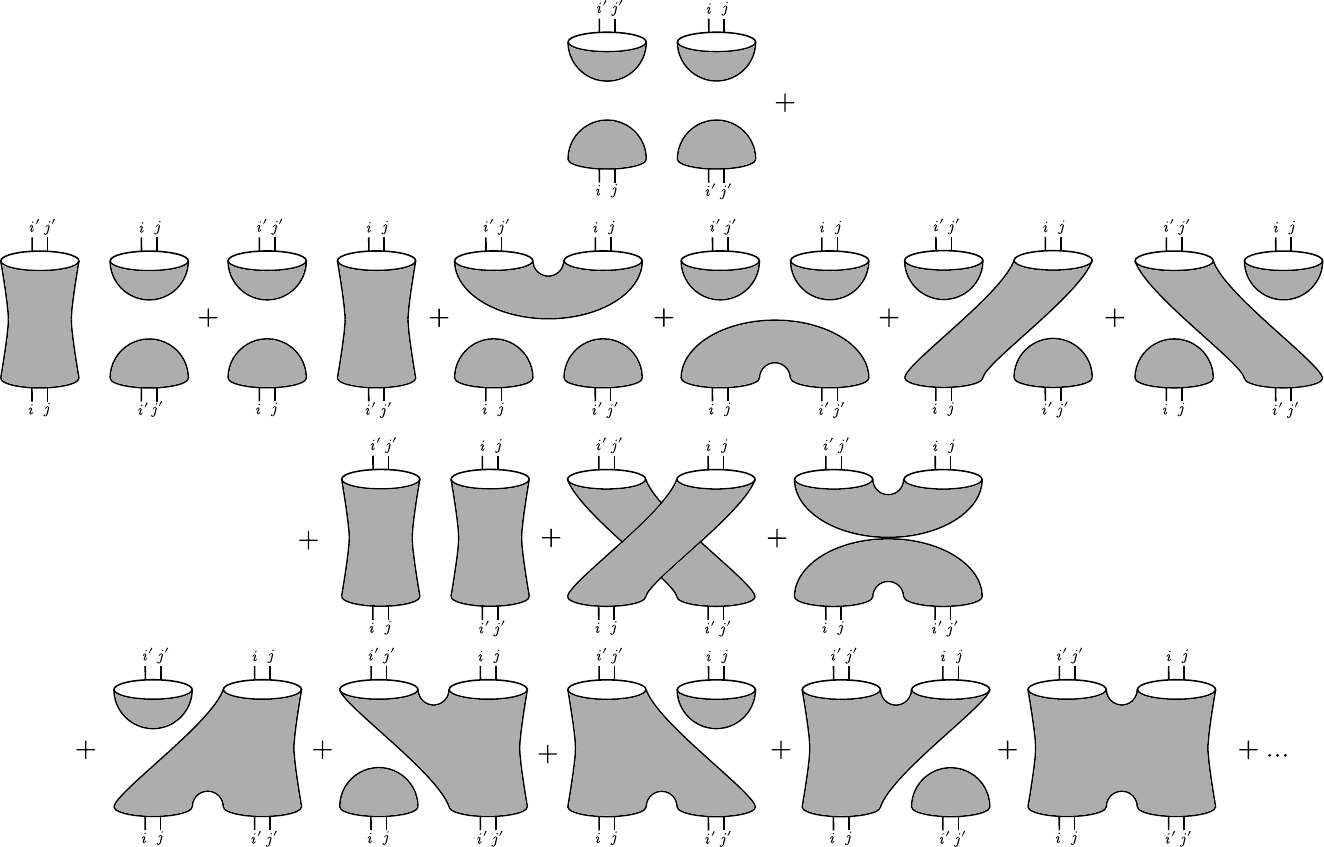}
\caption{Topologies contributing to the square of the two-particle one-universe inner product.  The ``$+\ldots$'' indicates summing over all the ways of adding handles to these topologies.}\label{2p2ipfig}
\efig
To compute the variance of the two-point function we need to compute the 15 topologies shown in figure \ref{2p2ipfig}.  To write out the answer in a compact form we can introduce the pairing symbol 
\be
\delta_{i_1\ldots i_{2n}}=\sum_{P}\prod_{(\ell,m)\in P}\delta_{i_\ell i_m},
\ee
which sums over all ways $P$ of pairing the $2n$ objects.  For example 
\be
\delta_{ijkm}=\delta_{ij}\delta_{km}+\delta_{ik}\delta_{jm}+\delta_{im}\delta_{jk}.
\ee
We then have the path integral result
\begin{align}\nonumber
\ol{|\llangle i'j'|ij\rrangle|^2}=&\frac{e^{4S_0}}{(1-e^{-2S_0})^4}\delta_{ij}\delta_{i'j'} +\frac{e^{2S_0}}{(1-e^{-2S_0})^3}\left(4\delta_{ij}\delta_{i'j'}\delta_{iji'j'}+\delta_{ij}\delta_{i'j'i'j'}+\delta_{i'j'}\delta_{ijij}\right)\\\nonumber
&+\frac{1}{(1-e^{-2S_0})^2}\left(\delta_{ijij}\delta_{i'j'i'j'}+2\delta_{iji'j'}\delta_{iji'j'}+2\delta_{ij}\delta_{iji'j'i'j'}+2\delta_{i'j'}\delta_{ijiji'j'}\right)\\
&+\frac{e^{-2S_0}}{1-e^{-2S_0}}\delta_{ijiji'j'i'j'}
\end{align}
Here the first line is the first two lines in figure \ref{2p2ipfig}, the second line is the third line and all but the last of the fourth line of figure \ref{2p2ipfig}, and the last line is the completely connected contribution.  We are supposed to subtract from this the absolute value squared of $\ol{\llangle i'j'|ij\rrangle}$, given by \eqref{2btop}, which gives variance
\begin{align}\nonumber
\ol{|\llangle i'j'|ij\rrangle-\ol{\llangle i'j'|ij\rrangle}|^2}=&\frac{e^{2S_0}}{(1-e^{-2S_0})^3}\left(2\delta_{ij}\delta_{i'j'}\delta_{iji'j'}+\delta_{ij}\delta_{i'j'i'j'}+\delta_{i'j'}\delta_{ijij}\right)\\\nonumber
&+\frac{1}{(1-e^{-2S_0})^2}\left(\delta_{ijij}\delta_{i'j'i'j'}+\delta_{iji'j'}\delta_{iji'j'}+2\delta_{ij}\delta_{iji'j'i'j'}+2\delta_{i'j'}\delta_{ijiji'j'}\right)\\
&+\frac{e^{-2S_0}}{1-e^{-2S_0}}\delta_{ijiji'j'i'j'}.
\end{align}
In order to decide whether or not this variance is small we should compare it to the product of the typical squared norms of the states, which is given by  
\be
\ol{\llangle ij|ij\rrangle}\,\,\ol{\llangle i'j'|i'j'\rrangle}=\left(\frac{e^{2S_0}}{(1-e^{-2S_0})^2}\delta_{ij}+\frac{1}{1-e^{-2S_0}}\delta_{ijij}\right)\left(\frac{e^{2S_0}}{(1-e^{-2S_0})^2}\delta_{i'j'}+\frac{1}{1-e^{-2S_0}}\delta_{i'j'i'j'}\right)
\ee
If $i=j$ and $i'=j'$ then the variance is of order $e^{2S_0}$ but product of squared norms is of order $e^{4S_0}$.  So again we see that the noise is exponentially suppressed relative to the signal, and thus
\be
\llangle i'i'|ii\rrangle \approx \aleph^{-1}[i'i'|ii].
\ee
On the other hand if $i\neq j$ then they both are either of order $e^{2S_0}$ if $i'=j'$ or of order $1$ if $i'\neq j'$.  Either way the noise is of order the signal, so the BUFT inner product has failed to emerge. Again this is quite similar to the spectral form factor, with the latter two situations being analogous to the late time behavior.   

\subsection{Including the observer}
In the previous section we saw that in general the equation \eqref{hbuqm} does not hold in a fixed holographic theory.  This is a generalization of the one-particle result \eqref{largevar1} to situations where there are disconnected contributions to the inner product.  In that context it was argued in \cite{Harlow:2025pvj} that the observer decoherence rule suppresses large fluctuations as in \eqref{obvar1}.  We now argue that the same is true in the two-particle model. 

\bfig
\includegraphics[height=6cm]{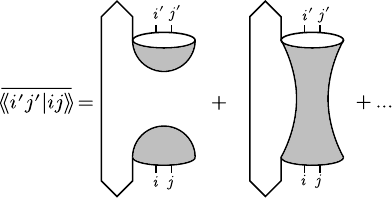}
\caption{Path integral contributions to the one-universe inner product in the presence of a decohered observer.  The observer line carries a species index with range $k_O$, which in both cases generates a factor of $k_O$ that is canceled by two factors of $1/\sqrt{k_O}$ from the normalization of the maximally mixed states.}\label{obsipfig}
\efig
Let's see what the observer rule does for our calculations.  We will consider an observer whose Hilbert space dimension is
\be
k_O=e^{S_{Ob}},
\ee
and maximally entangle them with an external clone $Ob'$.  The modified inner product is computed as in figure \ref{obsipfig}, the result is
\be\label{ipobs1p}
\ol{\lan i'j'|ij\ran}=\frac{e^{2S_0}}{(1-e^{-2S_0})^2}\lambda^2 \delta_{ij}\delta_{i'j'}+\frac{1}{1-e^{-2S_0}}(1+\lambda^2)\delta_{iji'j'}.
\ee
Here $\lambda$ is the amplitude to create an observer from the vacuum (meaning that an observer line can have an endpoint at the cost of a factor of $\lambda$.  We will assume that $\lambda$ is independent of the microstate of the observer.  We will work in the approximation that
\be
\lambda^2k_O\ll 1,
\ee
so that it is always preferable to have the observer connect between two boundary points rather than fluctuate even if this decreases the number of observer loops by one.  Thus factors of $\lambda$ will appear only when the geometry does not allow the observer to connect, as in the first geometry in figure \ref{obsipfig}.  We emphasize however that although this first term is suppressed by $\lambda^2$, it is \textit{enhanced} by $e^{2S_0}$.  We will focus on the situation where 
\be
k_O\ll e^{S_0}, 
\ee
in which case even in the presence of an observer the disconnected contribution to the transition amplitude can be dominant. 

\bfig
\includegraphics[height=14cm]{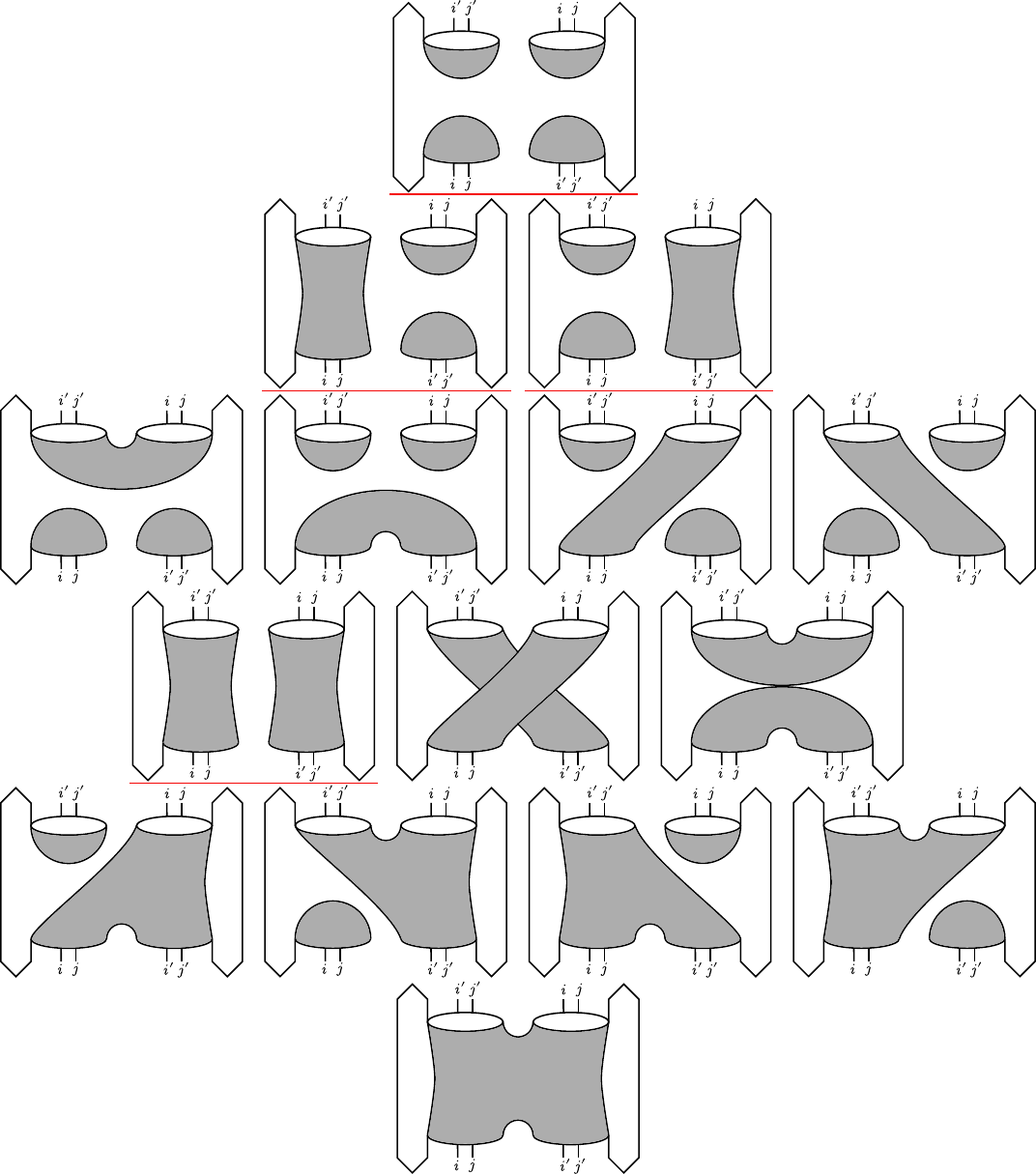}
\caption{Topologies contributing to $\overline{|\llangle i'j'|ij\rrangle|^2}$.  The topologies underlined in red are dominant at large $k_O$, and they also are precisely those appearing $|\overline{\llangle i'j'|ij\rrangle}|^2$.}\label{ip2obsfig}
\efig
To study the variance we need the average of the square of this inner product.  This has contributions from the fifteen topologies shown in figure \ref{2p2ipfig}, now modified to include the observer lines as in figure \ref{ip2obsfig}.  The full result is the following:
\begin{align}\nonumber
\overline{|\lan i'j'|ij\ran|^2}=&\frac{e^{4S_0}}{(1-e^{-2S_0})^4}\lambda^4 \delta_{ij}\delta_{i'j'}\\\nonumber
&+\frac{e^{2S_0}}{(1-e^{-2S_0})^3}\lambda^2\left[2\delta_{iji'j'}\delta_{ij}\delta_{i'j'}\left(1+\lambda^2 +(1+k_O\lambda^2)/k_O\right)+\left(\delta_{ijij}\delta_{i'j'}+\delta_{i'j'i'j'}\delta_{ij}\right)(1+k_O\lambda^2)/k_O  \right]\\\nonumber
&+\frac{1}{(1-e^{-2S_0})^2}\left[\delta_{ij,i'j'}\delta_{iji'j'}(1+\lambda^2)^2+\left(\delta_{ijij}\delta_{i'j'i'j'}+\delta_{iji'j'}\delta_{iji'j'}\right)(1+2\lambda^2+k_O\lambda^4)/k_O\right]\\\nonumber
&+\frac{2}{(1-e^{-2S_0})^2}\lambda^2(1+2/k_O+\lambda^2)\left[\delta_{ijiji'j'}\delta_{i'j'}+\delta_{iji'j'i'j'}\delta_{ij}\right]\\
&+\frac{e^{-2S_0}}{1-e^{-2S_0}}\delta_{ijiji'j'i'j'}\left((1+\lambda^2)^2+2(1+2\lambda^2)/k_O\right).
\end{align}
The first line is the first line of figure \ref{ip2obsfig}, the second line is the second and third lines of figure \ref{ip2obsfig}, the third line is the fourth line of figure \ref{ip2obsfig}, the fourth line is the fifth line of \ref{ip2obsfig}, and the fifth line is the sixth line of figure \ref{ip2obsfig}.  This result simplifies if we drop terms that are of order $1/k_{O}$ at each order in $e^{-S_0}$, after which we have
\begin{align}\nonumber
\overline{|\lan i'j'|ij\ran|^2}=&\frac{e^{4S_0}}{(1-e^{-2S_0})^4}\lambda^4 \delta_{ij}\delta_{i'j'}+\frac{2e^{2S_0}}{(1-e^{-2S_0})^3}\lambda^2\left[\delta_{iji'j'}\delta_{ij}\delta_{i'j'}+\ldots\right]\\
&+\frac{1}{(1-e^{-2S_0})^2}\left[\delta_{iji'j'}\delta_{iji'j'}+\ldots\right]+\frac{e^{-2S_0}}{1-e^{-2S_0}}\left[\delta_{ijiji'j'i'j'}+\ldots\right],
\end{align}
where $\ldots$ means terms which are of order $1/k_O$.  The key point is that at each order which is not non-perturbatively small in $S_0$, the leading term is precisely the same one we would get from the square of equation \eqref{ipobs1p}: the observer suppresses the difference between averaged holography and baby universe quantum mechanics!  More precisely we should compare the neglected terms to the product of squared norms 
\begin{align}\nonumber
\ol{\llangle ij|ij\rrangle}\, \ol{\llangle i'j'|i'j'\rrangle}=&\left(\frac{e^{2S_0}}{(1-e^{-2S_0})^2}\lambda^2 \delta_{ij}+\frac{1}{1-e^{-2S_0}}(1+\lambda^2)\delta_{ijij}\right)\\
&\times\left(\frac{e^{2S_0}}{(1-e^{-2S_0})^2}\lambda^2 \delta_{i'j'}+\frac{1}{1-e^{-2S_0}}(1+\lambda^2)\delta_{i'j'i'j'}\right).
\end{align}
When $i=j$ and $i'=j'$ the norm product is of order $\lambda^4e^{4S_0}$ while the variance is of order $\lambda^2 e^{2S_0}$, so the variance is small.  When $i\neq j$ but $i'=j'$ the norm squared is of order $\lambda^2e^{2S_0}$ and the variance is of order $\lambda^2 e^{2S_0}/k_O$, so the variance is now suppressed by a factor of $1/k_O$ compared to the previous subsection.  Finally when $i\neq j$ and $i'\neq j'$, the norm squared is of order $1$ and the variance is of order $1/k_O$.  Thus we see that the observer rule has succeeded in its task: in a fixed holographic theory the closed universe inner product coincides with the gravitational path integral result up to errors which are exponentially small in $S_{Ob}$!

\subsection{Comparing to canonical gravity}
We have now shown that using the observer rule, there is agreement \eqref{hbuqm} between holography and BUFT up to corrections of order $e^{-S_{Ob}/2}$.  What we have not shown is a similar agreement \eqref{hcqg} between either of these with CQG.  This is with good reason: the BUFT and CQG inner products are only close when the disconnected contribution is small enough that the cylinder dominates.  As we have seen, whether or not this happens is a state-dependent question.  For example in the two-particle topological model the cylinder dominates in $\aleph^{-1}[i'j'|ij]$ provided that $i\neq j$ or $i'\neq j'$.  More generally, in order for the cylinder to dominate a one-universe overlap $\aleph^{-1}[\phi|\psi]$, from equation \eqref{cqgHH} we need the product
\be
\lan\phi|HH\ran\lan HH|\psi\ran
\ee
of overlaps with the unnormalized canonical Hartle-Hawking state $|HH\ran$ to be small compared to $\lan \phi|\psi\ran$.  Since the norm of $|HH\ran$ is $\log \aleph$, this means that that we need at least one of the overlaps with the normalized canonical HH state 
\be
|\wt{HH}\ran=\frac{|HH\ran}{\sqrt{\log \aleph}}
\ee
to be substantially smaller than $\frac{1}{\sqrt{\log \aleph}}$.  For example in dS space we have
\be
\log \aleph \sim e^{S_{dS}},
\ee
where $S_{dS}$ is the Gibbons-Hawking entropy of de Sitter space,
so we need the overlap with $|\wt{HH}\ran$ to be smaller than it would be for a randomly chosen state in a Hilbert space whose dimension is $e^{S_{dS}}$.  On the other hand for closed universes with negative cosmological constant it is quite natural for $\lan HH|\phi\ran$ to be small, for example if we take the spatial topology to be a round $\mathbb{S}^{d-1}$ then a natural saddle point for $\lan HH|\phi\ran$ would be the Poincare ball but this does not have any spherical totally geodesic surface that we could use to glue onto a Lorentzian bang/crunch solution with a semiclassical description.  Since we presumably are interested in states which do have a semiclassical Lorentzian interpretation, it is plausible that the sources we use to create these will also result in an inner product that is dominated by the cylinder.  For more discussion of the dominance of connected vs disconnected saddles in this context see \cite{Marolf:2021kjc}.  

\section{Patch operators}\label{patchsec}
We have now shown that BUFT can emerge from holography in a fixed theory up to errors of order $e^{-S_{Ob}/2}$ provided that we use the observer rule of \cite{Harlow:2025pvj}.  We have also seen however that this emergent BUFT only agrees with CQG in states where the BUFT inner product is dominated by the cylinder.  What are we to make of situations where the disconnected contribution dominates?  In such situations the transition amplitudes are dominated by a process where the initial state of the entire universe ``unfluctuates'' and then the final state ``refluctuates'' out of nothing.  This does not sound like a process that would be pleasant to experience, but it is hard to tell without coming up with some more precise way of asking about the experiences of observers in a situation where the initial state of the universe is essentially irrelevant.  In this final section we will describe a possible approach to doing this.  The basic idea is that we will try to reformulate the above discussion in terms of what we will call \textbf{patch operators}, which roughly speaking are integrals of local operators over all of spacetime.  More precisely we define them to include a conditioning on having an observer present at some spacetime point that we integrate over, and allowing the observer to make measurements in their local vicinity of operators whose locations are defined relative to that observer (such operators were recently advocated in \cite{Chandrasekaran:2022cip,Witten:2023xze}).  We will see that in situations where the HH norm $\aleph$ is large such operators have expectation values that are close to their CQG values in the HH state regardless of the choice of global state.

It must be acknowledged that there are very serious challenges to having a fluctuation-driven universe be phenomenologically viable, essentially because of the infamous ``Boltzmann brain'' problem.  In this paper we do not have much to say about this, as we are focused on just defining some mathematical theory of cosmology, but this is of course an important problem to return to in the future.

The treatment of patch operators in this section is somewhat cursory, they will be discussed in much more detail in the companion paper by Ying Zhao.  

\subsection{Operators in BUFT}
We begin by recalling how we developed BUFT in section \ref{pisec}.  Namely we used the gravitational path integral to define a hermitian product on the linear span of the states $|J_1\ldots J_N]$ via equation \eqref{BUFTIP}, and then, assuming this hermitian product is positive semidefinite, we quotiented by the set of null states to get the BUFT Hilbert space \cite{Marolf:2020xie}.  

We can similarly try using the path integral to define operators on this Hilbert space.  For example given any local scalar operator $X^{loc}$ we can define a diffeomorphism-invariant operator $X$ by integrating $X^{loc}$ over spacetime:
\be\label{Xdef}
[J_1'\ldots J_N'|X|J_1\ldots J_M]\equiv\sum_{\M} \int \mathcal{D}\phi \left(\int_{\M} d^d x X^{loc}(x)\right)e^{-S_E[\phi,\M]}.
\ee 
Here $\M$ are all manifolds obeying the boundary conditions determined by $J_1,J_2,\ldots$ and $J_1',J_2',\ldots$.  If we divide by $\aleph$ then we get a sum only over manifolds where each connected component contains a boundary and/or $X^{loc}$.  To the extent that this defines a valid operator $X$, it must be one that commutes with $\hat{Z}(J)$,
\be
[\J'|X\hat{Z}(J)|\J]=[\J'|\hat{Z}(J)X|\J],
\ee
since the left and right sides of this equation are defined by path integrals with the same boundary conditions.  It therefore will be diagonal in the basis of $|\alpha]$ states.  Here we have again introduced the abbreviated notation $\J$ to indicate some number of $J$ sources.

To be sure that the quantity $[\J'|X|\J]$ computed by the path integral actually defines a good operator $X$ on the span of the $|\J]$ states however, a necessary and sufficient condition is that for any null state
\be
|\omega]=\int d\J \omega(\J)|\J],
\ee
we have
\be\label{omega1}
[\omega |X|\J]\equiv\int d\J'\omega(\J')^*[\J'|X|\J]=0
\ee
for all $\J$.  Moreover if we want this operator to respect the quotient by null states, we further require that
\be\label{omega2}
[\J'|X|\omega]\equiv \int d\J\omega(\J)[\J'|X|\J]=0.
\ee
Demonstrating these conditions directly from the path integral is difficult, which is perhaps why operators were not discussed in \cite{Marolf:2020xie}. On the other hand the quantity \eqref{Xdef} is clearly well-defined and diffeomorphism invariant, and it naturally generalizes what we would call the matrix elements of an operator in the ordinary quantum field theory path integral.  Our attitude is that the conditions \eqref{omega1} and \eqref{omega2} are somewhat analogous to the positive semi-definiteness of the hermitian product \eqref{BUFTIP}: they are not manifest starting from a generic path integral, but they had better be true in any path integral that emerges from holography.  In fact we can actually show that this is the case, analogously to our demonstration of positive semidefiniteness  at the end of section \ref{avsubsec}.  The idea is that since any patch operator $X$ commutes with $\hat{Z}(J)$, we can specify it by giving its eigenvalues $\wt{X}_\alpha$ in the $|\alpha]$ basis:
\be
[\J'|X|\J]=\sum_\alpha \wt{X}_\alpha [\J'|\alpha][\alpha|\J].
\ee
These coefficients automatically obeys \eqref{omega1} and \eqref{omega2} since the $|\alpha]$ states are orthogonal to any null states.  

\subsection{Examples of patch operator calculations}
Let's now get a little practice using patch operators.  We will work at the level of pictures rather than equations in concrete models, since that will be enough to make the points we wish to make, but we will use the topological model in $1+1$ to get a sense of the size of the contributions from various topologies.  

\bfig
\includegraphics[height=3cm]{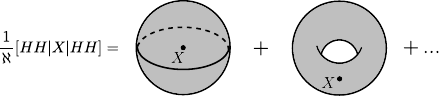}
\caption{Computing the expectation value of a patch operator in the Hartle-Hawking state.}\label{XHHfig}
\efig
The first quantity we can consider is the expectation value of a single patch operator in the Hartle-Hawking state.  The result is shown in figure \ref{XHHfig}.  In $1+1$ dimensions this calculation is dominated by the sphere provided that $X$ is not tuned to make its sphere expectation value small compared to its natural value of $e^{2S_0}$, with higher genus contributions suppressed by powers of $e^{-2S_0}$.  Similarly in de Sitter space in $3+1$ dimensions the sphere is the dominant contribution.  

\bfig
\includegraphics[height=6cm]{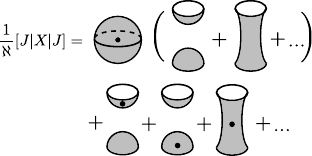}
\caption{Computing the expectation value of a patch operator in another one-universe state.  When the sphere expectation value is large this calculation is dominated by the first line, whose $J$-dependence cancels out when we normalize the expectation value.}\label{HHJJfig}
\efig
The second quantity we will consider is the expectation value of a patch operator in a single-universe state $|J]$.  The calculation is shown in figure \ref{HHJJfig}.  The key point is then that when the sphere expectation value is large then
\be
\frac{[J|X|J]}{[J|J]}\approx \aleph^{-1}[HH|X|HH],
\ee
so in other words the choice of initial state has almost no effect on the normalized expectation value of $X$!  This is an example of the fluctuation dominance discussed at the beginning of the section.  On the other hand if the sphere is small or zero, and the same is true for the hemispheres, then the cylinder is the dominant contribution and we recover the CQG expectation value in the state $|J\ran$.

\bfig
\includegraphics[height=2.5cm]{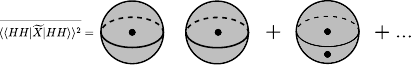}
\caption{Computing the variance of the Hartle-Hawking expectation value of a patch operator.  The first term cancels when we subtract the square of the expectation value, so the variance is controlled by a single sphere.  When the sphere expectation value is large the variance is small compared to the squared expectation value.}\label{Xvarfig}
\efig
Finally we can consider how the fixed theory answer $\wt{X}$ compares to its ensemble average, to get a sense of to what extent the path integral answer emerges from a definite holographic dual with no $\alpha$ parameters.  This calculation is shown in figure \ref{Xvarfig}, the main point is that the dominant contribution to the variance is of order $e^{2S_0}$, but this is much smaller than the square of the expectation value, which is of order $e^{4S_0}$.  Thus the gravitational answer emerges nicely in a fixed holographic dual, without need for any average over $\alpha$ parameters (or any implementation of the observer rule!)  Again the same conclusion will hold whenever the sphere partition function is large, as it is for example in de Sitter space.

\subsection{Patch operators in the two-particle closed universe code}
\bfig
\includegraphics[height=7cm]{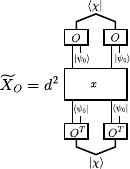}
\caption{Encoding a patch operator $X$ into the holographic code for the two-particle model.  Here $x$ is the same operator on the canonical one-universe Hilbert space.}\label{Xcodefig}
\efig
In this section we briefly explain how to include patch operators into the two-particle holographic code model from section \ref{codesec}.  The idea again is to give a direct prescription for computing $\wt{X}_\alpha$ starting from some CQG operator $x$.  The rule is shown in figure \ref{Xcodefig}.  For example using this rule in the HH state, we have
\be
\ol{\llangle HH|\wt{X}_\alpha|HH\rrangle}=\int dO \wt{X}_\alpha=\lan HH|X|HH\ran+\Tr\left((1+S)x\right),
\ee
where $S$ is the swap operator on the two particles and $|HH\ran$ is the canonical Hartle Hawking state given by equation \eqref{2pHH}.  The first term here matches the sphere contribution in figure \ref{XHHfig} and the second term matches the torus.  

\subsection{Patch operators in the landscape}
As a final application of patch operators, we will show that they can lead to somewhat puzzling conclusions in the context of a landscape with multiple vacua such as is expected to exist in string theory \cite{Susskind:2003kw}.  We will consider a toy landscape with only two minima, a stable AdS minimum and a metastable dS minimum.  The only asymptotic spatial or Euclidean boundaries this theory allows are AdS bouundaries, and in particular we should expect the Hilbert space of this theory to decompose into sectors based on the spatial AdS boundaries that are present.  By AdS/CFT the Hilbert space in each sector is just the Hilbert space of the dual CFT on the appropriate spatial manifold.  We will consider a state $|\psi]$ in the sector with one AdS boundary present whose spatial manifold is a round sphere, and in particular we will take this state to contain an observer sitting in AdS (to sharpen the paradox we will assume that both vacua allow for observers).  For example the observer could be created by a source at earlier times on the boundary.  We emphasize that the state $|\psi]$ is a quite conventional state in AdS/CFT, its energy is small compared the the black hole threshold and the usual dictionary would tell us that the dual spacetime is AdS with an observer sitting in it.  

\bfig
\includegraphics[height=4cm]{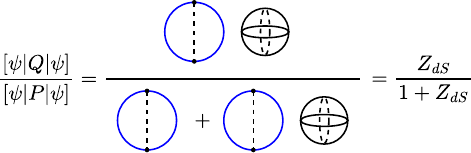}
\caption{Using patch operators to compute the probability of being in dS or AdS in a landscape with two vacua.  The AdS boundary is shaded blue, while the dS sphere is shown in black.  Dashed lines indicate observers - the AdS region always has one because we put it there using a boundary source, while the dS region can have one via a fluctuation in the HH state.}\label{dsadsfig}
\efig
The idea however is to use patch operators to decide in this state whether an observer should expect to find themself in AdS or dS.  We will consider two patch operators:
\begin{align}
Q&=\begin{cases} 0 & \mathrm{no\,\, observer}\\ 0 &\mathrm{observer \,\, in \,\, AdS}\\ 1 & \mathrm{observer \,\, in \,\, dS}
\end{cases} \\
P&=\begin{cases} 0 & \mathrm{no\,\, observer}\\ 1 & \mathrm{observer \,\, in \,\, dS\,\,or\,\,AdS}
\end{cases}
\end{align}
Each of these operators is integrated throughout spacetime, performing the indicated tests locally.  The ratio
\be
p_{dS}=\frac{[\psi|Q|\psi]}{[\psi|P|\psi]}
\ee
gives the probability that the observer is in dS.  The calculation of $p_{dS}$ is shown in figure \ref{dsadsfig}.  In order for the patch operator in the numerator to give a nonzero answer it must be located in the dS region, while in the denominator the patch operator can be in either region so there are two terms.  The AdS contribution cancels between the numerator and the denominator, so the result depends only on the conditioned de Sitter sphere partition function, which should approximately be given by
\be
Z_{dS}\approx e^{S_{dS}-m_{Ob}/T_{dS}}.
\ee
The mass of a human-sized observer in units of the de Sitter temperature in our universe is of order $10^{70}$, but this is far smaller than the de Sitter entropy which is  of order $10^{120}$.  Thus we conclude that
\be
p_{dS}=1-O(e^{-S_{dS}}),
\ee
so the observer almost surely finds themself in de Sitter space even though we started with a CFT state whose bulk dual is viewed as quite standard!  At least in this theory, the Boltzmann brain problem has come to invade AdS.  Clearly this conclusion requires substantially more thought than we have given it, but for now we will leave things here.  

\paragraph{Acknowledgments:}  I thank Netta Engelhardt, Luca Iliesiu, Adam Levine, Hong Liu, Juan Maldacena, Arvin Shahbazi-Moghaddam, Douglas Stanford, Lenny Susskind, Herman Verlinde, and Edward Witten for helpful discussions. This work was carried out in discussion and collaboration with Ying Zhao, to whom I am grateful for many conversations.  I am supported by the Packard Foundation as a Packard Fellow, the US Department of Energy under grants DE-SC0012567 and DE-SC0020360, and the MIT department of physics.

\bibliographystyle{jhep}
\bibliography{bibliography}

\end{document}